\documentstyle[12pt,epsfig,cite]{article}

\renewcommand{\today}{9 January 1995\\
Revised: 6 January 1997}

\newcommand{\nc}{\newcommand}
\nc{\be}{\begin{equation}}
\nc{\ee}{\end{equation}}
\nc{\bea}{\begin{eqnarray}}
\nc{\eea}{\end{eqnarray}}
\nc{\beas}{\begin{eqnarray*}}
\nc{\eeas}{\end{eqnarray*}}
\nc{\noi}{\noindent}
\nc{\etal}{{\it et al.}}
\nc{\sD}{\not \! \! D}
\nc{\s}[1]{\not \! #1}
\nc{\non}{\nonumber}
\nc{\bb}{\bibitem}
\nc{\rw}{$\rho\!-\!\omega$ }
\nc{\lf}{\left}
\nc{\r}{\right}
\nc{\mb}[1]{\makebox[#1]{}}
\nc{\pa}{\partial}
\nc{\sA}{\not \! \! A}
\nc{\newsec}[1]{\section{#1}\mb{0.5cm}}
\nc{\h}{\frac{1}{2}}
\nc{\ra}{\rightarrow}
\nc{\la}{\leftarrow}
\nc{\ep}{$e^+e^-\ra\pi^+\pi^-$}
\def\mathunderaccent#1{\let\theaccent#1\mathpalette\putaccentunder}
\def\putaccentunder#1#2{\oalign{$#1#2$\crcr\hidewidth
\vbox to.2ex{\hbox{$#1\theaccent{}$}\vss}\hidewidth}}

\nc{\ti}{\mathunderaccent\tilde}

\nc{\M}{{\cal M}}

\begin{document}
\thispagestyle{empty}
\begin{flushright}
ADP-95-1/T168 \\
hep-ph/9501251
\end{flushright}

\begin{center}
{\large{\bf Rho-omega mixing, vector meson dominance \\
and the pion
form-factor}} \\
\vspace{2.2 cm}
H.B.~O'Connell, B.C.~Pearce, \\
A.W.~Thomas and A.G.~Williams \\
\vspace{1.2 cm}
{\it
Department of Physics and Mathematical Physics \\
University of Adelaide, S.Aust 5005, Australia } \\
\vspace{1.2 cm}
\today
\vspace{1.2 cm}
\begin{abstract}

We review the current status of \rw mixing and 
discuss its implication for our understanding of charge-symmetry
breaking. In order to place this work in context we also review the
photon-hadron coupling within the framework of vector meson dominance.
This leads naturally to a discussion of the electromagnetic form-factor
of the pion and of nuclear shadowing.

\end{abstract}

\vspace{2.5cm}
{ Published in Prog. Nucl. Part. Phys. {\bf 39} (1997) 201-252.}\\
(Edited by Amand Faessler.)\\
\end{center}

\newpage
\tableofcontents
\newpage
\listoffigures

\newpage
\section{Introduction}
\mb{.5cm} 

Charge symmetry is broken at the most fundamental level in strong
interaction physics through the small mass difference between up and
down quarks in the QCD Lagrangian. As a consequence the physical $\rho$
and $\omega$ mesons are not eigenstates of isospin but, for example, the
physical $\rho$ contains a small admixture of an $I = 0$ $q\bar{q}$
state. This phenomenon, known loosely as \rw  mixing, has been observed
in the charge form-factor of the pion, which is dominated by the $\rho$
in the time-like region. Indeed, vector meson dominance (VMD) was
constructed to take advantage of this fact.

Nuclear physics involves strongly interacting systems which are not yet
amenable to calculations based directly on QCD itself. Instead the
nucleon-nucleon (NN) force is often treated in a semi-phenomenological
manner using a one- or (two-) boson exchange model. Within such a
framework, \rw mixing gives rise to a charge symmetry violating (CSV) NN
potential which has been remarkably effective in explaining measured CSV
in nuclear systems -- notably in connection with the
Okamoto-Nolen-Schiffer anomaly in mirror nuclei. However, the
theoretical consistency of this approach has been challenged by recent
work suggesting that the \rw mixing amplitude changes sign between
the $\rho$ pole and the space-like region involved in the NN
interaction.

Our aim is to provide a clear, up-to-date account of the ideas of VMD as
they relate particularly to the pion form-factor and to \rw mixing.
We begin with an historical review of VMD in Sec.~2. The evidence for
\rw mixing at the $\rho$ pole is presented in Sec.~3 along with the
standard theoretical treatment. In Sec.~4 we briefly highlight the
role played by \rw mixing in the traditional formulation of the CSV NN
force. More modern theoretical concerns about the theoretical
consistency of the usual approach are summarised in Sec.~5, while in
Sec.~6 these new ideas are tested against the form-factor data. In
Sec.~7 we make a few remarks concerning shadowing in the light of our
new appreciation of VMD,  summarise our conclusions
and outline some open problems.

\section{Vector Meson Dominance}

\mb{.5cm} The physics of hadrons was a topic of intense study
long before the gauge field
theory of quantum chromodynamics (QCD)
now believed to describe it completely
was invented. Hadronic physics was described using a variety of models and
incorporating approximate symmetries.
It is a testimony to the insight behind these models (and the inherent
difficulties in solving
non-perturbative QCD) that they still play an important role in our 
understanding.

\begin{figure}[htb]
  \centering{\
     \epsfig{angle=270,figure=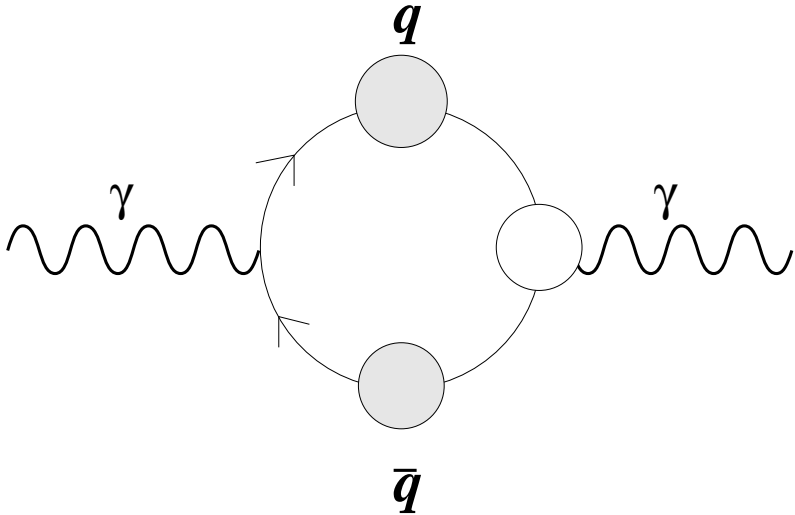,height=7cm}  }
\parbox{130mm}{\caption{One-particle-irreducible QCD contribution to the photon propagator.}
\label{qloop}}
\end{figure}

One particularly important aspect of hadronic physics which concerns us here
is the interaction between the
photon and hadronic matter \cite{We}. This has been remarkably well 
described using
the vector meson dominance (VMD) model. This assumes that the hadronic
components of the vacuum polarisation of the photon consist exclusively of the
known vector mesons. This is certainly an approximation, but in the regions
around the vector meson masses, it appears to be a very good one.
As vector mesons are believed to be bound states of
quark-antiquark pairs
\cite{Frank,Frank2,Frank3}, it is tempting to try to establish a connection 
between the old language of VMD and the Standard Model \cite{ba}. In the 
Standard Model, quarks, being charged, couple to 
the photon and so the strong sector  contribution
to the photon propagator arises, in a manner
analogous to the electron-positron loops
in QED, as shown in Fig.~\ref{qloop}.

The diagram contains dressed quark propagators and the proper
(i.e., one-particle irreducible)
photon-quark vertex (the shaded circles include
one-particle-reducible parts, while the empty circles are
one-particle-irreducible \cite{RW}). In QED 
we can approximate the photon self-energy reasonably well using
bare propagators and
vertices without worrying about higher-order dressing.
However, in QCD, the dressing of these quark loops 
can not be so readily dismissed as being of higher order in a perturbative
expansion. (Although for the heavier quarks, higher order effects
can be ignored as a consequence of asymptotic freedom \cite{GW}, 
one must be careful about this \cite{Craig}.)
No direct translation between
the Standard Model and VMD has yet been made.

\subsection{Historical development of VMD}
\mb{.5cm}
The seeds of VMD were sown by Nambu \cite{Nam} in 1957 when he suggested 
that the charge distribution of the proton and neutron, as determined by
electron scattering, could be accounted for by a heavy neutral vector meson
contributing to the nucleon form factor. This isospin-zero field
is now called the $\omega$.

The anomalous magnetic moment of the nucleon was believed to be dominated
by a two-pion state \cite{CKGZ}. The pion form-factor, $F_\pi(q^2)$,
(to be discussed later in some detail) was taken to be unity
in these initial calculations ---i.e., the pions were treated as point-like
objects. 
By 1959 Frazer and
Fulco \cite{FF} concluded (after an investigation of analytic
structure) that the pion form-factor had
to satisfy the dispersion relation
\be
F_\pi(q^2)=1+\frac{q^2}{\pi}\int_{4m_\pi^2}dr\frac{{\cal I}m\; F_\pi(r)}{
r(r-q^2-i\epsilon)}
\label{f&f}
\ee
and that to be consistent with data a suitable peak in the pion form-factor
was required, which they believed could result from a strong pion-pion
interaction.
The analytic structure of the partial wave amplitude in the
physical region could be approximated as a pole of appropriate position
and residue (a successful approximation in nucleon-nucleon scattering). An
analysis determined that the residue should be positive, raising the 
possibility of a resonance, which we now know as the $\rho^0$.

It was Sakurai who proposed a theory of the strong interaction
mediated by vector mesons \cite{Sak} based on the non-Abelian field theory of 
Yang and Mills \cite{YM}. He was deeply troubled by the problem of the
masses of the mesons in such a theory, as they would destroy the local
(flavour) gauge invariance. He published his
work with this matter unresolved in the hope that it would stimulate further
interest in the field.

\begin{figure}[htb]
  \centering{\
     \epsfig{angle=270,figure=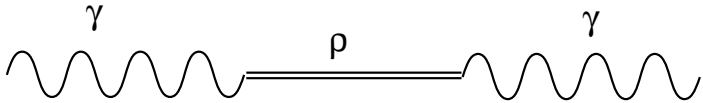,height=3cm}  }
\parbox{130mm}{\caption{A simple VMD-picture 
representation of the hadronic contribution to the photon propagator.
The heavier vector mesons are included in
generalised VMD models.}
\label{vdmpol}}
\end{figure}

Kroll, Lee and Zumino did pursue the idea of reproducing VMD from  field
theory \cite{KLZ}.
Within the simplest VMD model the hadronic contribution 
to the polarisation of the photon takes
the form of a propagating vector meson (see Fig.~\ref{vdmpol}).
This now replaces the QCD contribution to the 
polarisation process depicted in Fig.~\ref{qloop}.

This form
arises from the assumption that the hadronic electromagnetic current operator,
$j_\mu^{\rm EM}$,
is proportional to the field operators of the vector mesons (multiplied by
their mass squared). This is referred to as the  the field-current identity.  
This is then included in
the general structure of the hadronic part of the Lagrangian, giving a precise
formulation of VMD in terms of a local, Lagrangian field theory.
One starts with the identity for the neutral $\rho$-meson 
\be
[j^{\rm EM}_\mu(x)]_{I=1}=\frac{m_\rho^2}{g_\rho}\rho_\mu^0(x),
\label{fci}
\ee
and then generalises \cite{LZ} to an isovector field, ${\vec \rho}(x)$, of
which $\rho^0(x)$ is the third component [i.e., $\rho^0(x)\equiv \rho^3(x)$]. 
Eq.~(\ref{fci}) implies that the 
field ${\vec \rho}(x)$ is divergenceless under the strong interaction,
which is just the usual Proca condition
\be
\pa_\mu \vec{\rho}^{\;\mu}=0,
\label{proca}
\ee
for a massive vector field coupling to a conserved current.
The resulting Lagrangian for the hadronic sector is the same as
the (flavour) Yang-Mills Lagrangian \cite{YM}, but also has a mass 
term which destroys
the local gauge invariance. 
Although gauge invariance is necessary for 
renormalisability\footnote{In general
there are only two cases in which a massive vector field is renormalisable, 
see Ref.~\cite{CL}, p.~61:\\
a) a gauge theory with mass generated by spontaneous symmetry breaking; \\
b) a theory with a massive vector boson coupled to a conserved
current and without additional self-interactions.}, 
Kroll {\it et al.}
were unconcerned by this; stating that the non-zero value for
the mass made it possible to connect the field conservation
equation, Eq.~(\ref{proca}), with the equation
of motion of the field. The case of a global SU(2) massive vector 
field (the $\rho$-field)
interacting with a triplet pion field and
coupled to a conserved current is treated in detail by Lurie \cite{Lu}.

\subsection{Gauge invariance and VMD}
\mb{.5cm}

Sakurai's analysis of VMD \cite{Sak2,HFN} takes place in the context of a local
gauge theory. Although a mass term in the Lagrangian breaks gauge symmetry,
Sakurai viewed the generation of interactions by minimal substitution in
the Lagrangian to be interesting enough to ignore this problem. Lurie \cite{Lu}
has discussed the $\rho,\,\pi,\,N$ system
using coupling to conserved currents which reproduces
Sakurai's results. As it only assumes the Lagrangian to be invariant under
{\em global} SU(2), the appearance of mass terms causes no difficulty.
One can then examine how to include the photon in this system.
Lurie's primary concern was to have the $\rho$ couple to a conserved current,
and he did this by constructing a Lagrangian whose equation of motion had the 
Noether current associated with the global
SU(2) symmetry appearing on the right hand side. In doing this, he arrives 
at the standard non-Abelian Lagrangian (given in p.~700 of Ref.~\cite{IZ}),
which is where we start.

We begin with the Lagrangian (while Sakurai and Lurie worked in a
Euclidean metric, we follow the conventions of Bjorken and Drell \cite{BjD})
\be
{\cal L}_{\rm full}=-\frac{1}{4} \vec{\rho}_{\mu\nu}\cdot \vec{\rho}^{\;\mu\nu}
+\h m_\rho^2\vec{\rho}_{\mu}\cdot\vec{\rho}^{\mu}+
\h D_\mu\vec{\pi}\cdot D^\mu\vec{\pi}
-\h m_\pi^2\vec{\pi}\cdot\vec{\pi},
\label{rholag}
\ee
where
\be
\vec{\rho}_{\mu\nu}=\pa_\mu \vec{\rho}_\nu-\pa_\nu \vec{\rho}_\mu
-g\vec{\rho}_{\mu}\times\vec{\rho}_{\nu},
\label{fieldstrength}
\ee
and\footnote{
We use hermitian T's given by the algebra $[T^a,T^b]=-ic^{abc}T^c$
and normalised by Tr$(T^aT^b)=\delta_{ab}/2$. Thus, in the adjoint 
representation, $(T^c)_{ba}=-ic^{cab}$.
\label{Tdefn}
}
\bea
D_\mu\vec{\pi}&=&(\pa_\mu-ig\vec{\rho}_\mu\cdot\vec{T})\vec{\pi},
\label{covderiv} \\
&=& \pa_\mu\vec{\pi}-g\vec{\rho}\times\vec{\pi}.
\label{covdiv}
\eea
This Lagrangian is symmetric under the transformation
\be
\vec{\phi}\ra\vec{\phi}+\vec{\phi}\times\vec{\epsilon},
\ee
where $\vec{\phi}$ represents the isovector fields 
of the $\vec{\rho}$ and $\vec{\pi}$.
The generation of interactions from minimal substitution is used by Sakurai
and Lurie to motivate universality (i.e., the coupling constant of the
$\rho$
introduced via the covariant derivative, $D_\mu$,
is the same for all particles). However, as a slight violation to this
rule is seen experimentally, we shall distinguish between $g$
and the constant $g_{\rho}$ appearing in Eq.~(\ref{fci}), which Sakurai equates
in order to satisfy a constraint on the pion form-factor (to be
discussed later).

{}From Eq.~(\ref{covdiv}) it follows that
\be
\h D_\mu\vec{\pi}\cdot D^\mu\vec{\pi}
=\h \pa_\mu\vec{\pi}\cdot \pa^\mu\vec{\pi}
-g\vec{\rho}_\mu\cdot(\vec{\pi}\times\pa^\mu\vec{\pi})+
\h g^2(\vec{\rho}_\mu\times\vec{\pi})^2.
\label{sea}
\ee
After some algebra we obtain the equation of motion for the $\rho$ field
\be
\pa_\nu\vec{\rho}^{\;\nu\mu}+m_\rho^2\vec{\rho}^{\;\mu}
=g\vec{J}^\mu_{\rm Noether}
\label{noetherform}
\ee
where the Noether current is 
\be
\vec{J}^\mu_{\rm Noether}=-\frac{\pa {\cal L}}{\pa(\pa_\mu\vec{\rho}_\nu)}
\times\vec{\rho}_\nu-\frac{\pa {\cal L}}{\pa(\pa_\mu\vec{\pi})}\times\vec{\pi}
\ee
giving
\be
\vec{J}^\mu_{\rm Noether}= \vec{\rho}^{\;\mu\nu}\times\vec{\rho}_\nu+
\vec{\pi}\times\pa^\mu\vec{\pi}
+g(\vec{\rho}^{\;\mu}\times\vec{\pi})\times\vec{\pi}.
\label{ncurrent}
\ee
As the Noether current is necessarily conserved, Eq.~(\ref{noetherform})
tells us that the field is divergenceless, as in Eq.~(\ref{proca}). 
Transferring the non-Abelian part of the field strength tensor (the
cross product in Eq.~(\ref{fieldstrength})) to the RHS of 
Eq.~(\ref{noetherform}) gives us,
\be
\pa_\nu(\pa^\nu\vec{\rho}^{\;\mu}-\pa^\mu\vec{\rho}^{\;\nu})
+m_\rho^2\vec{\rho}^{\;\mu}=g(\vec{J}^\mu_{\rm Noether}+
\pa_\nu(\vec{\rho}^{\;\nu}\times\vec{\rho}^{\;\mu})).
\ee
Again using the fact that the $\rho$ field is divergenceless 
(Eq.~(\ref{proca})),
we can rewrite the equation of motion in the inverse propagator form
\be
(\pa^2+m_\rho^2)\vec{\rho}^{\;\mu}=g\vec{J}^{\mu},
\label{eqmot}
\ee
where $\vec{J}_\mu$ is also a divergenceless current given by
\bea
\nonumber
\vec{J}^\mu &=&\vec{J}^\mu_{\rm Noether}+
\pa_\nu(\vec{\rho}^{\;\nu}\times\vec{\rho}^{\;\mu}) \\
&=&\vec{J}^\mu_{\rm Noether}+
\vec{\rho}^{\;\nu}\times\pa_\nu\vec{\rho}^{\;\mu}.
\label{current}
\eea
As Lurie notes, the presence of the $\rho$ field itself in 
$\vec{J}^\mu_{\rm Noether}$  
prevents us from writing the interaction part of the Lagrangian in the simple
$\vec{\rho}_\mu\cdot \vec{J}^\mu$ fashion (which 
is possible for the fermion-vector interaction). 
A similar situation  for scalar electrodynamics is discussed
by Itzykson and Zuber \cite{IZ} (p.~31--33).

\begin{figure}[htb]
  \centering{\
     \epsfig{angle=270,figure=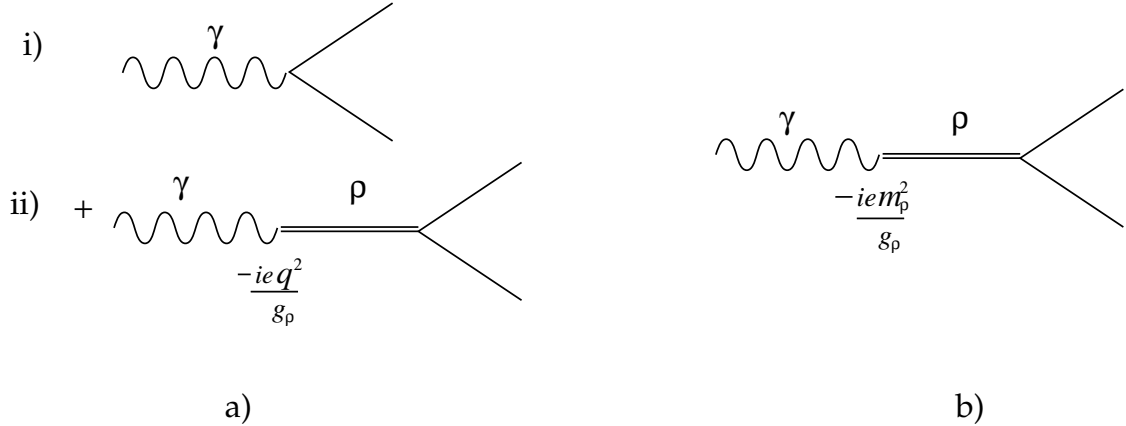,height=5.5cm}  }
\parbox{130mm}{\caption{Contributions to the 
pion form-factor in the two representations of vector meson dominance
a) VMD1 b) VMD2.}
\label{twoways}}
\end{figure}

Our task now is to include electromagnetism in this model, and to do this
we shall allow Eq.~(\ref{fci}) to guide us.
Eqs.~(\ref{fci}) and (\ref{eqmot}) imply (as $\pa_\mu \ra iq_\mu$) a 
corresponding matrix element relation for the electromagnetic 
interaction\footnote{We take $e$ to be positive, $e=|e|$.}
\bea
\nonumber
\left<B\right|ej^{\rm EM}_\mu\left|A\right>&=&
e\left<B\right|\frac{m_\rho^2}{g_\rho}\rho_\mu^3\left|A\right> \\
&=&e\frac{m_\rho^2}{g_\rho}
\left<B\right|\frac{-gJ_\mu^3}{q^2-m_\rho^2}\left|A\right> \\
&=&\frac{-iem_\rho^2}{g_\rho}\frac{-i}{q^2-m_\rho^2}
\left<B\right|gJ_\mu^3\left|A\right>.
\label{msquared}
\eea
This is to say that the photon appears to couple to the hadronic field via
a $\rho$ meson, to which it couples with strength $em_\rho^2/g_\rho$.
(This model is illustrated in Fig.~\ref{twoways}b, below.)

Before proceeding, we shall make, as Sakurai does, the simplifying assumption
that one can neglect the $\rho$ self-interaction (from
now on we shall refer only to the $\rho^0\equiv\rho^3$),
i.e., the parts of the current given by Eq.~(\ref{current}) involving
$\rho$ terms,
and concern ourselves only with the piece of the current that looks like
\be
J^\mu_\pi=(\vec{\pi}\times\pa^\mu\vec{\pi})_0,
\label{piJ}
\ee
which we shall refer to now simply as $J^\mu$. Changing from a Cartesian
to a charge basis, we can re-write Eq.~(\ref{piJ}) as
\be
J_\mu=i(\pi^-\pa_\mu\pi^+-\pi^+\pa_\mu\pi^-).
\label{picurrent}
\ee
As the $\rho^0$ decays almost entirely via the two-pion channel, this is a
reasonable approximation for the current. 
We can then write the simple linear coupling term 
in the Lagrangian, and we shall choose to write $g$ as $g_{\rho\pi\pi}$
\be
{\cal L}_{\rho\pi}=-g_{\rho\pi\pi}\rho_\mu J^\mu.
\ee

\begin{figure}[htb]
  \centering{\
     \epsfig{angle=270,figure=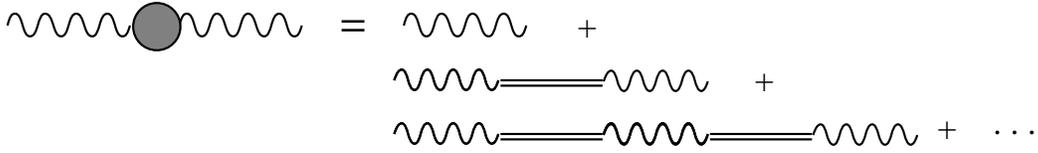,height=2cm}  }
\parbox{130mm}{\caption{VMD dressing of the photon propagator by a series
of intermediate $\rho$ propagators.}
\label{modprop}}
\end{figure}

The important problem now is to ensure that after adding electromagnetism 
we still have a gauge invariant theory.
The naive $\gamma\!-\!\rho$ vertex prescription usually seen in discussions
of VMD,
\[
-\frac{em_\rho^2}{g_\rho},
\]
as motivated by Eq.~(\ref{msquared}),
suggests a coupling term in the effective Lagrangian of the form 
\be
{\cal L}_{\rm eff}=-\frac{em_\rho^2}{g_\rho}\rho^3_\mu A^\mu.
\label{efflang}
\ee
This is suggested by the substitution of the field current identity 
(Eq.~(\ref{fci})) into the interaction piece of the electromagnetic Lagrangian,
$-ej^{\rm EM}_\mu A^\mu$. However electromagnetism cannot be 
incorporated into 
Eq.~(\ref{rholag}) simply by adding Eq.~(\ref{efflang})
and a kinetic term for the photon. This would result
in the photon acquiring an {\em imaginary} mass \cite{Sak} when one considers
the dressing of the photon propagator in the manner of Fig.~\ref{modprop}
using $\rho-\gamma$ vertices determined by Eq.~(\ref{efflang}).

However, we can find a term that emulates Eq.~(\ref{efflang}), but ensures
that the photon remains massless.
Such a term is
\be
{\cal L}_{\gamma\rho}=-\frac{e}{2g_\rho}F_{\mu\nu}\rho^{\mu\nu}.
\label{gammarho}
\ee
We need to re-express this in momentum space which can be done using
integration by parts to transform $\pa_\mu A_\nu
\pa^\mu \rho^\nu$ to $-\pa_\mu\pa^\mu A_\nu\rho^\nu$ and then send
$\pa_\mu\ra iq_\mu$ giving 
\be
F_{\mu\nu}\rho^{\mu\nu}\ra 2q^2A_\mu\rho^\mu.
\label{qsquared}
\ee
The other term in $F_{\mu\nu}\rho^{\mu\nu}$ 
can be discarded because it contains a piece
that can be written as $q_\mu \rho^\mu$ and thus vanishes as the $\rho$
field is divergenceless. 

However, the interaction Lagrangian of Eq.~(\ref{gammarho})
is not sufficient as it would decouple the photon from the $\rho$
(and hence then from hadronic matter) at $q^2=0$. What is needed is another
term which directly couples the photon to hadronic matter. This is
\be
-eA_\mu J^\mu,
\label{l2}
\ee
where $J_\mu$ is the hadronic current to which the $\rho$ couples,
the pion component of which is
given in Eq.~(\ref{piJ}).
Thus we have an interaction between the photon and hadronic matter of 
exactly the same
form as that between the $\rho$ and hadronic matter (though 
suppressed by a factor of $e/g_{\rho\pi\pi}$). 
This term is most noticeable at $q^2=0$ where the influence of
the $\rho$-meson in the photon-pion interaction vanishes. 

To summarise the arguments just given, the 
photon and vector meson part of the Lagrangian we require is
\be
{\cal L}_{\rm VMD1}=-\frac{1}{4}F_{\mu\nu}F^{\mu\nu}-\frac{1}{4}
\rho_{\mu\nu}\rho^{\mu\nu}+\h m_\rho^2\rho_\mu\rho^\mu-
g_{\rho\pi\pi}\rho_\mu J^\mu 
-eA_\mu J^\mu-\frac{e}{2g_\rho} F_{\mu\nu}\rho^{\mu\nu}.
\label{vmdlag}
\ee
We shall refer to this as the {\em first representation} of VMD. We note that
this representation has a direct photon---matter coupling as well
as a photon---$\rho$ coupling which vanishes at $q^2=0$.

Sakurai also outlined an alternative formulation of VMD, which has survived
to become the standard representation. In many ways it is not as elegant as
the first; for instance, the Lagrangian has a photon mass term.
Despite this it has established itself
as the most popular representation of VMD:
\be
{\cal L}_{\rm VMD2}= -\frac{1}{4}(F'_{\mu\nu})^2-\frac{1}{4}
(\rho'_{\mu\nu})^2+\h m_\rho^2(\rho'_\mu)^2
-g_{\rho\pi\pi}\rho'_\mu J^\mu-\frac{e' m_\rho^2}{g_\rho}\rho'_\mu
A'^\mu+\h\left(\frac{e'}{g_\rho}\right)^2 m_\rho^2 (A'_\mu)^2.
\label{newvmd}
\ee
In the limit of universality 
($g_\rho\!=\!g_{\rho\pi\pi}$) the two representations become equivalent
and one can transform between them using
\bea
\rho'_\mu &=& \rho_\mu+\frac{e}{g_\rho}A_\mu, \label{transfs1} \\
A'_\mu &=&A_\mu \sqrt{1-\left(\frac{e}{g_\rho}\right)^2}, \\
e' &=& e\sqrt{1-\left(\frac{e}{g_\rho}\right)^2}.
\label{transfs}
\eea
Substituting for $\rho'_\mu,A'_\mu$ and $e'$ in Eq.~(\ref{newvmd})
gives Eq.~(\ref{vmdlag}) $+O((e/g_\rho)^3)$.
We shall refer to Eq.~(\ref{newvmd}) as the {\em second representation} of VMD.

The appearance of a photon mass term at first seems slightly troublesome.
However, when dressing the photon in the manner of Fig.~\ref{modprop}, we
see that the propagator has the correct form as $q^2\ra 0$. We have
\be
iD(q^2)=
\frac{-i}{q^2-\frac{e^2m_\rho^2}{g_\rho^2}}+
\frac{-i}{q^2-\frac{e^2m_\rho^2}{g_\rho^2}}\frac{-iem_\rho^2}{g_\rho}
\frac{-i}{q^2-m_\rho^2}\frac{-iem_\rho^2}{g_\rho}
\frac{-i}{q^2-\frac{e^2m_\rho^2}{g_\rho^2}} +\cdots
\ee
Summing this using the general operator identity
\be
\frac{1}{A-B}=\frac{1}{A}+\frac{1}{A}B\frac{1}{A}+
\frac{1}{A}B\frac{1}{A}B\frac{1}{A}+\cdots
\label{opidentity}
\ee
we obtain ($m\equiv m_\rho$)
\bea
\nonumber
iD(q^2)&=&-i
\left[q^2-\frac{e^2m^2}{g_\rho^2}-\frac{e^2m^4}{g_\rho^2(q^2-m^2)}\right]^{-1}
\\
&=&-i
\left[q^2-\frac{e^2m^2}{g_\rho^2}+\right.
\left.\frac{e^2m^2}{g_\rho^2(1-{q^2}/{m^2})}\right]^{-1}
\label{modifprop} \\
&\ra&\frac{-i}{q^2(1+e^2/g_\rho^2)}
\eea
as $q^2\ra 0$. We are thus left with a modification to the coupling constant 
\be
e^2\ra e^2(1-e^2/g^2_\rho),
\ee
and interestingly we see that the photon propagator is significantly
modified away from $q^2=0$.

We conclude this discussion 
with a comparison of the use of the two models by describing
the process $\gamma\ra\pi^+\pi^-$.
We can identify the relevant terms in the Lagrangian for each case. 
{}From
${\cal L}_{\rm VMD1}$ (Eq.~(\ref{vmdlag})) and ${\cal L}_{\rm VMD2}$ 
(Eq.~(\ref{newvmd})) we have, respectively,
\bea
{\cal L}_{1}&=&-\frac{e}{2g_\rho}F_{\mu\nu}\rho^{\mu\nu}
-eJ_\mu A^\mu-g_{\rho\pi\pi} \rho^\mu J_\mu, 
\label{vmd1} \\
{\cal L}_{2}&=&-\frac{em_\rho^2}{g_\rho}\rho_\mu A^\mu-
g_{\rho\pi\pi}\rho_\mu J^\mu.
\label{vmd2}
\eea

If the photon coupled to the pions directly,
then the Feynman amplitude for this process would be (as in scalar 
electrodynamics \cite{IZ})
\be
{\cal M}^{\mu}_{\gamma\rightarrow\pi^{+}\pi^{-}} = 
\langle\pi^+\pi^-|eJ^\mu |0\rangle =-e(p^+-p^-)^\mu,
\label{sed}
\ee
Where $J_{\mu}$ is given in Eq.~(\ref{picurrent}).  
However, in the presence of the vector
meson interactions of Eqs.~(\ref{vmd1}) and (\ref{vmd2}), the total amplitude
is modified.  The pion form factor, $F_\pi(q^2)$, which represents the 
contribution from the intermediate
steps connecting the photon to the pions, is defined by the relation
\be
{\cal M}^{\mu}_{\gamma\rightarrow\pi^{+}\pi^{-}} = -e(p^+-p^-)^\mu
F_\pi(q^2),
\label{ffdef}
\ee
where now ${\cal M}^{\mu}_{\gamma\rightarrow\pi^{+}\pi^{-}}$ is the full
amplitude including all possible processes.
The form-factor is the multiplicative deviation from a pointlike
behaviour of the coupling of the photon to the pion field.  We discuss
$F_\pi(q^2)$ in detail later.

To lowest order, 
we have for ${\cal L}_1$ (see Eq.~(\ref{qsquared}))
\be
F_\pi(q^2)=
\left[1-\frac{q^2}{q^2-m_\rho^2}\frac{g_{\rho\pi\pi}}{g_{\rho}}\right],
\label{E1}
\ee
and for ${\cal L}_2$
\be
F_\pi(q^2)=-
\frac{m_\rho^2}{q^2-m_\rho^2}\frac{g_{\rho\pi\pi}}{g_{\rho}}.
\label{E2}
\ee
In the limit of zero momentum transfer, the photon ``sees" only the charge
of the pions, and hence we must have
\be
F_\pi(0)=1.
\label{ffzero}
\ee
The reader may notice that Eq.~(\ref{ffzero}) is automatically satisfied by the
dispersion relation of Frazer and Fulco, Eq.~(\ref{f&f}) and by 
VMD1 (Eq.~(\ref{E1}))
but must be imposed on the VMD2 result (Eq.~(\ref{E2})) by demanding
$g_{\rho\pi\pi}=g_{\rho}$. 

This is the basis of Sakurai's argument for
universality mentioned earlier, i.e., that the photon couples to the $\rho$
as in Eq.~(\ref{vmd2}) and that therefore $g_{\rho\pi\pi}$ must
equal $g_{\rho}$. This is a direct consequence of assuming
{\em complete} $\rho$ dominance of the form-factor
(i.e., VMD2). The second part of universality, namely that
$g_{\rho\pi\pi}=g_{\rho NN}=...=g_\rho$ results from the assumption
that the interactions are all generated from the gauge principle (i.e., by
minimal substitution for the covariant derivative given in 
Eq.~(\ref{covderiv})).

As Sakurai points out, the two representations
of VMD are equivalent
in the limit of universality (as we would expect from
Eqs.~(\ref{transfs1}--\ref{transfs})). Without universality {\em only} 
VMD1, maintains the condition $F_\pi(0)=1$.
Due to the popularity of the second interpretation, though, $F_\pi(0)=1$ is 
more often viewed as a constraint on various introduced parameters \cite{BCP}.
We illustrate the difference between the
two representations in Fig.~\ref{twoways}.

Interestingly,
Caldi and Pagels \cite{CP} arrived at a similar expression for the 
pion form-factor to Eq.~(\ref{E1}) from a {\em direct}
photon contribution {\em and} a fixed $\gamma\!-\!\rho$ vertex. Their coupling
of the $\rho$ to the pion field, though, is momentum dependent, and it is
because of this that they reproduced the first representation.

\subsection{The $\rho$ as a dynamical gauge boson}
\mb{.5cm}
Bando {\it et al.} have succeeded in
constructing a local gauge model which reproduces
VMD \cite{Ban}. This model is based on the idea of a hidden local symmetry
originally developed in supergravity theories. The $\rho$-meson appears 
as the {\em dynamical} gauge boson of a {\em hidden} local symmetry 
in the non-linear, chiral Lagrangian. The mass of the $\rho$ is
generated by the Higgs mechanism associated with the hidden local 
symmetry.

We begin with the Lagrangian of the non-linear sigma-model \cite{Bha}
\be
{\cal L}=\frac{f_\pi^2}{4}{\rm Tr}[\pa_\mu U\pa^\mu U^{\dagger}],
\ee
where $f_\pi$ is the pion decay constant (93 MeV) and
\be
U(x)=\exp[2i \pi(x)/f_\pi].
\ee
Here $\pi(x)$ are the pion fields $\pi(x)=\pi^aT^a$, where $T^a$ are the
generators of SU(2) (see footnote \ref{Tdefn}
on page \pageref{Tdefn}). The field $U$
transforms under chiral ${\rm SU}(2)_L\otimes{\rm SU}(2)_R$ as:
\be
U(x)\ra g_L U(x)g^{\dagger}_R,
\label{Ueq}
\ee
where $g_{L,R}\in {\rm SU}(2)_{L,R}.$

As it stands this particular Lagrangian is invariant under 
global ${\rm SU}(2)_L\otimes{\rm SU}(2)_R$. 
However, it can be cast into a form which possesses, in
addition, a local (and hidden) ${\rm SU}(2)_V$ symmetry. 
We can separate $U(x)$ into two constituents which transform respectively
under left and right ${\rm SU}(2)$
\be
U(x)\equiv \xi_L^{\dagger}(x)\xi_R(x)
\ee
where the $\xi(x)$ are SU(2) matrix-valued entities transforming like
\be
\xi_{L,R}\ra \xi_{L,R}g^{\dagger}_{L,R}.
\ee
However, the interesting part comes in supposing these components
also possess a local ${\rm SU}(2)_V$ symmetry,
\be
\xi_{L,R}\ra h(x)\xi_{L,R},
\label{gauge}
\ee
where $h(x)=e^{-i\vec{\alpha}(x)\cdot\vec{T}}$.
The important point here is that the field $U(x)$
does not ``see'' this local ${\rm SU}(2)_V$ transformation (because it is
invariant under it, even though its components $\xi$ are not), and thus we
say it is a {\em hidden symmetry}.

The ${\rm SU}(2)_V$ invariant Lagrangian can now be re-written \cite{Bha} as
\be 
{\cal L}=f^2_\pi {\rm Tr}\left[\frac{1}{2i}( \pa_\mu\xi_L\xi^{\dagger}_L
-\pa_\mu\xi_R\xi_R^{\dagger})\right]^2.
\label{bandolag1}
\ee
However, if we now introduce a gauge field
\[
V_\mu=\vec{V}_\mu\cdot\vec{T},
\]
and covariant derivative (c.f. Eq.~(\ref{covdiv}))
\be
D_\mu \xi_{L,R} =\pa_\mu \xi_{L,R}-ig V_\mu \xi_{L,R}.
\ee
We can write the original Lagrangian as \cite{Bha}
\be
{\cal L}_1 = f^2_\pi {\rm Tr}\left[\frac{1}{2i}(D_\mu\xi_L\xi^{\dagger}_L
-D_\mu\xi_R\xi_R^{\dagger})\right]^2,
\label{lag1}
\ee
which is easily seen to revert to Eq.~(\ref{bandolag1}) upon substitution
for the covariant derivatives.
We now similarly construct
\bea
{\cal L}_2 &=& f^2_\pi  {\rm Tr}\left[\frac{1}{2i}(D_\mu\xi_L\xi^{\dagger}_L
+D_\mu\xi_R\xi_R^{\dagger})\right]^2 \\
&=&g^2 f^2_\pi {\rm Tr}\left[V_\mu-\frac{1}{2ig}(\pa_\mu\xi_L\xi^{\dagger}_L
+\pa_\mu\xi_R\xi_R^{\dagger})\right]^2
\eea
which is invariant under the local ${\rm SU}(2)_V$ transformation $h(x)$
provided that $V_\mu$ transforms under ${\rm SU}(2)_V$ as
\be 
V \ra h(x)Vh^{\dagger}(x)+\frac{i}{g}\;h(x)\pa_\mu h^{\dagger}(x).
\ee
Interestingly, the Euler-Lagrange equation for $V_\mu$ is 
\be
\frac{\pa {\cal L}}{\pa V_\mu}-\pa_\nu\left(\right.\frac{\pa {\cal L}}{\pa
(\pa_\nu V_\mu)}\left.\right)=0,
\ee
which implies that ${\cal L}_2=0$. Thus 
we need to do something to enable us to keep
our vector field, $V_\mu(x)$.  Bando {\it et al.} assumed that
quantum (or {\em dynamical}) effects at the ``composite level" (where the
underlying quark
substructure brings QCD into play) generate the kinetic term of the
gauge field $V_\mu(x)$
\[
-\frac{1}{4}\vec{F}_{\mu\nu}\cdot\vec{F}^{\mu\nu}
\]
where, like $\vec{\rho}^{\;\mu\nu}$ in Eq.~(\ref{fieldstrength}),
\be
\vec{F}_{\mu\nu}=\pa_\mu \vec{V}_\nu-\pa_\nu \vec{V}_\mu
-g\vec{V}_\mu\times\vec{V}_\nu.
\ee
{}From this we construct a new Lagrangian of the form 
\be
{\cal L}={\cal L}_1+a{\cal L}_2 -\frac{1}{4}\vec{F}_{\mu\nu}
\cdot\vec{F}^{\mu\nu},
\ee
where $a$ is an arbitrary parameter. We now fix the ${\rm SU}(2)_V$ gauge
(Eq.~(\ref{gauge})) by imposing the condition
\be
\xi_L^{\dagger}(x)=\xi_R(x)=\xi(x)=e^{i\vec{\pi}(x)\cdot\vec{T}/f_\pi}.
\label{gfix}
\ee
Approximating $\xi$ by $(1+i\vec{\pi}\cdot\vec{T}/f_\pi)$ our Lagrangian
now has the form\footnote{For SU(2) $\{T^a,T^b\}=\delta_{ab}/2$ hence 
$T^aT^b=-i\epsilon^{abc}T^c/2+\delta_{ab}/4$.}
\be
{\cal L}
=\frac{f_\pi^2}{4}{\rm Tr}[\pa_\mu U\pa^\mu U^{\dagger}]
+\h a g^2 f_\pi^2\vec{V}_\mu \cdot\vec{V}^{\mu}-\h a g
\vec{V}\cdot\vec{\pi}\times 
\pa_\mu\vec{\pi}-\frac{1}{4}\vec{F}_{\mu\nu}\cdot\vec{F}^{\mu\nu},
\ee
to order $\pi^2$.
We can identify, by comparison with Eq.~(\ref{rholag}),
\be
m_\rho^2=a g^2 f_\pi^2, 
\ee
and from Eq.~(\ref{ncurrent}) we recognise the current
\be
\vec{J}^{V}_\mu= \vec{\pi}\times \pa_\mu\vec{\pi},
\ee
and hence,
\be
g_{V\pi\pi}=\h a g.
\ee

The next step towards reproducing VMD is to incorporate electromagnetism.
We extend the hidden gauge group to a larger group, 
${\rm SU}(2)_V\otimes{\rm U}(1)_Q$ where ${\rm U}(1)_Q$ is {\em not}
a hidden symmetry as
\be
U\ra b(x)Ub^{\dagger}(x).
\label{btransf}
\ee
The transformation $b(x)\in {\rm U}(1)_Q=\exp(-i e_0 Q \theta(x))$, where
$Q$ is the generator of the one-parameter U(1) group (analogous to $T^a$
for SU($N$)). The EM field couples to
\be
Q=\h Y + T_3,
\ee
where $Y$ is the hypercharge, which is zero in this case. Bando {\it et al.}
draw attention to the complete independence of the $\rho$ and photon source
charges, which produces a simple picture.

The transformation given in Eq.~(\ref{btransf}) means that the $\xi$ fields
transform like
\be
\xi_{L,R}\ra\xi_{L,R}b^{\dagger}.
\ee
We therefore require for a covariant derivative,
\be
D_\mu\xi_{L,R}=\pa_\mu\xi_{L,R}-ig\vec{V}_\mu \vec{T}
\xi_{L,R}-ie_{0}\xi_{L,R}B_\mu T^3,
\label{emderiv}
\ee
where $B_\mu$ is essentially the photon field.
With this we find that the relevant parts of the Lagrangian, namely
\[
D_\mu\xi_{L}\xi_{L}^{\dagger}\pm D_\mu\xi_{R}\xi_{R}^{\dagger} 
\]
are invariant under ${\rm U(1)_Q}$, provided $B_\mu$ transforms like
\be
B_\mu\ra B_\mu-\frac{i}{e_{0}}\;\pa_\mu b^{\dagger} b.
\ee
Incorporating our new covariant derivative (Eq.~(\ref{emderiv})) we have
as the new Lagrangian 
\be
{\cal L}={\cal L}_1+a{\cal L}_2
-\frac{1}{4}\vec{F}_{\mu\nu}\cdot\vec{F}^{\mu\nu} 
-\frac{1}{4}B_{\mu\nu}B^{\mu\nu},
\ee
where $B_{\mu\nu}$ is the strength tensor of the field $B_\mu$.
We now once again fix the gauge in the manner of Eq.~(\ref{gfix}).

Expanding $\xi$ once again to first order in $\vec{\pi}$, the new
Lagrangian becomes
\bea
\nonumber
{\cal L}={ \frac{f_\pi^2}{4}}{\rm Tr}[\pa_\mu U\pa^\mu U^{\dagger}]
-{ \frac{1}{4}}\vec{F}_{\mu\nu}\cdot\vec{F}^{\mu\nu}-
{ \frac{1}{4}}(\pa_\mu B_\nu-\pa_\nu B_\mu)^2 \\
\nonumber
+{ \h} m_\rho^2 \vec{V}_\mu\cdot\vec{V}^\mu-
{ \frac{1}{g}}
e_0 m_\rho^2
V_\mu^3 B^\mu+{ \h }(\frac{e_0}{g})^2m_\rho^2 B^\mu B_\mu \\
-{ \h} a g \vec{V}_\mu\cdot(\pa^\mu\vec{\pi}\times\vec{\pi})
-e_0(1-{ \frac{a}{2}})B^\mu (\pa^\mu\vec{\pi}\times\vec{\pi})_3.
\label{bandolag}
\eea
We are now free to choose a value for $a$.
Choosing $a=2$ both reproduces the VMD2 Lagrangian
given in Eq.~(\ref{newvmd}) and imposes universality as 
$g_{\rho\pi\pi}=ag/2=g_\rho$. One would then be free to make the
transformations given by Eq.~(\ref{transfs}) to obtain 
VMD1 (Eq.~(\ref{vmdlag})).

However, instead of doing this
Bando {\it et al.} follow the procedure for removing the mass of the U(1) field
in the Standard Model \cite{ba}, where an almost identical
situation occurs for the photon and the $Z^0$.
One says that the states $V_3^\mu$ and $B^\mu$ mix,
spontaneously breaking the ${\rm SU}(2)_V\otimes{\rm U}(1)_Q$ down to
${\rm U}(1)_{\rm em}$. We set, as opposed to 
Eqs.~(\ref{transfs1})-(\ref{transfs}),
\bea
A_\mu=\frac{1}{\sqrt{g^2+e_0^2}}(g B_\mu+e_0 V^3_\mu) \\
\label{bando1}
V^0_\mu=\frac{1}{\sqrt{g^2+e_0^2}}(gV^3_\mu-e_0B_\mu)
\label{bando2}
\eea
and the photon mass vanishes as required. 
The relevant part of the resulting Lagrangian is now
\bea
{\cal L}=-\frac{1}{4}(A_{\mu\nu}A^{\mu\nu}+V^0_{\mu\nu}V^{0{\mu\nu}})
+\h (m_{\rho^0})^2V^0_\mu V^{0\mu}-(g_{\rho\pi\pi}V^0_\mu+eA_\mu)\epsilon_{3ab}
\pa^\mu \pi^a\pi^b,
\label{lastbando}
\eea
where
\beas
m_{\rho^0}^2&=&a(g^2+e_0^2)f_\pi^2, \\
e&=&\frac{e_0g}{\sqrt{g^2+e_0^2}}, \\
g_{\rho\pi\pi}&=&\frac{ge}{e_0},
\eeas
for $a\!=\!2$.

We note that this Lagrangian has no explicit
coupling between the photon and the $\rho^0$, although
there is a direct coupling of the photon to the hadronic current.
They can mix, however,
via a pion-loop, which results in a $q^2$ dependent mixing between the photon
and the $\rho$-meson. Because we are working to lowest
order in the pion field Eq.~(\ref{lastbando})
lacks the seagull term
(i.e., one of the form of the final term in Eq.~(\ref{sea}))
the resulting mixing amplitude 
will be neither transverse, nor vanish at $q^2=0$ (see section \ref{rwzero}). 
However, this is a departure from the usual formulations of VMD which
contain an explicit mixing term in the Lagrangian.
Bhaduri merely notes that once this transformation is made
the physical photon now has a hadron-like part through Eq.~(\ref{bando1})
\cite{Bha}. This issue is analysed in more detail by Schechter
\cite{Schecter}. He considers the diagonal basis to be
the physical one (as the photon is massless and
gauge invariance is preserved) and argues that the
vector meson supplies a $q^2$ correction to the pion form
factor, rather than giving the whole thing.

Hung has extended this model to include the weak bosons \cite{Hung}. What is
especially interesting about his work in light of our presentation is
his reproduction of the first representation of VMD, which he demonstrates
is equivalent to ``precisely the old vector meson dominance'' (by which
he means the second representation), as universality is a consequence
of his model.

\subsection{Summary}
\mb{.5cm}
We have described how the interactions of the photon
with hadronic systems can be
usefully modelled using vector mesons. This idea was then moulded into 
a Lagrangian field theory, but the masses of the vector meson prevented
one from having a gauge invariant theory. Two equivalent formulations
of VMD were developed, VMD1 in which the coupling of the photon to the 
$\rho$ is momentum dependent 
(vanishing at $q^2=0$), and VMD2 where it is not. 
If universality is imposed
these representations produce the same physics.

In an attempt to put VMD on a more solid theoretical footing, Bando 
{\it et al.} were able to write down
a gauge invariant theory which reduces (c.f. Eq.~(\ref{bandolag})) to the VMD2
Lagrangian when one expands to second order in the pion field.

A unified picture of these above mentioned phenomenological
approaches is afforded by the bosonised Nambu--Jona-Lasinio (NJL) model
\cite{ER}. The NJL model features a four-point quark interaction. Bosonising
this chirally invariant model automatically yields the field current
identity of Eq.~(\ref{fci}) and, from this, VMD. The bosonised
NJL model also contains the hidden local symmetry of Bando {\it et al.}
and one can demonstrate, through a chiral
rotation, that the two effective meson Lagrangians are equivalent.
In the full
quantum theory, the two representations, VMD1 and VMD2 are related
by a simple change of mesonic integration variables \cite{reinhardt}.

\section{\rw mixing}
\mb{.5cm}
We shall discuss here how \rw mixing was seen experimentally and 
the challenge it presented to physicists to explain the
mechanism driving it. The importance of \rw mixing 
in the conventional understanding of charge symmetry violation (CSV)
in nuclear physics (c.f.~Sec.~\ref{nucl})
has made it crucial for us to improve our understanding of this phenomenon.

\subsection{The electromagnetic form-factor of the pion}
\mb{.5cm}
One problem in which VMD found particular success was the description 
of the electromagnetic form-factor of the pion \cite{GS}. 
As this has played such a crucial role in our understanding of \rw mixing
it is useful to outline what we mean by it and how the theoretical
predictions are compared with experimental data.

\begin{figure}[htb]
  \centering{\
     \epsfig{angle=270,figure=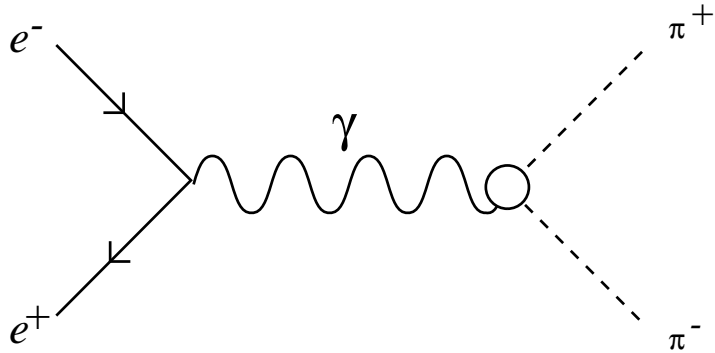,height=5cm}  }
\parbox{130mm}{\caption{Electron-positron pair annihilating
to form a photon which then decays to a pion pair.}
\label{fig:s}}
\end{figure}

We are 
concerned with the $s$-channel process depicted in Fig.~\ref{fig:s}, 
in which an electron-positron pair
annihilate, forming a photon which then decays to two pions.
We define the form-factor, $F_\pi(s)$, by Eq.~(\ref{ffdef}). The form-factor 
represents all possible strong
interactions occurring within the circle in Fig.~\ref{fig:s}, which we
model using VMD.

In the time-like region, $F_\pi(q^2)$ is measured experimentally
in the process $e^+e^-\ra\pi^+\pi^-$, which, to lowest order in $e^2$,
is given by the process shown in  Fig.~\ref{fig:s}. The momenta
of the electron and positron are $p_1$ and $p_2$ respectively, and
$p_3$ and $p_4$ are the momenta of the $\pi^+$ and $\pi^-$.
The differential cross-section is given by
\be
\frac{d\sigma}{d\Omega}=\frac{\vec{p}_3^{\;2}}{|\vec{p}_3|(p^0_3+p^0_4)-p^0_3
\hat{p}_2\cdot(\vec{p}_1+\vec{p}_2)}\frac{\frac{1}{4}\Sigma_{\rm pols}|
{\cal M}_{fi}|^2}{64 \pi^2 \sqrt{(p_1\cdot p_2)^2-m_e^4}},
\ee
where $\hat{p}$ is the unit vector in the direction of $\vec{p}$.
We are thus interested in calculating the Feynman amplitude, ${\cal M}_{fi}$, 
for this process. The 
leptonic and photon part of the diagram are completely 
standard. The interesting part of the diagram concerns the coupling of 
the photon to the pion pair
represented by Fig.~\ref{fig:s}. The form of this part 
of the diagram, 
${\cal M}_{\gamma\ra\pi^+\pi^-}$, is given in Eq.~(\ref{ffdef}).
In full, the amplitude is
\be
{\cal M}_{fi}=\overline{v}(2)ie\gamma^\mu u(1)iD_{\mu \nu}(q)
eF_\pi(q^2)(p_4-p_3)^\nu,
\label{eq:m}
\ee
with the photon propagator being given by 
\be
iD_{\mu \nu}(q)=\frac{(-i)}{q^2}\left[ g_{\mu \nu}+(\xi -1)\frac{q_\mu 
q_\nu}{q^2}\right].
\label{photonprop}
\ee
Particular choices of $\xi$ correspond to 
particular covariant gauges. The second term in Eq.~(\ref{photonprop}) vanishes
because the phton couples to conserved currents.

In the centre of mass frame in which we set
$|\vec{p}|=p$, we have $E^2-p^2=m_e^2$, 
$E^2-p'^2=m_\pi^2$, and 
$\vec{p}\cdot\vec{p}\;'=-pp'\cos\theta$.
Using $\sqrt{s}=2E$ the differential cross-section becomes
\be
\frac{d\sigma}{d\Omega}=\frac{e^4}{s^2}\frac{(\frac{1}{4}s-m_\pi^2)^{1/2}}{(
\frac{1}{4}s^2-s m_e^2)^{1/2}}\frac{(E^4-E^2 m_\pi^2-((E^4-E^2 (m_\pi^2+m_e^2)
+m_\pi^2m_e^2)\cos^2\theta)) |F_\pi(s)|^2}{\sqrt{s}\; 64 \pi^2}.
\ee
Since we have $m_e^2<<m_\pi^2<s$, we can simplify the above formula to
\[
\frac{d\sigma}{d\Omega}=\frac{e^4}{s^2}\frac{(s-4m_\pi^2)^{1/2}}{s\sqrt{s}}
\frac{1}{8\pi^2}(E^4-E^2 m_\pi^2)(1-\cos^2\theta) |F_\pi(s)|^2.
\]
{}From this we obtain the total cross-section 
\be
\sigma=\frac{\alpha^2\pi}{3}\frac{(s-4m_\pi^2)^{3/2}}{s^{5/2}}|F_\pi(s)|^2.
\label{xsection}
\ee

\begin{figure}[htb]
  \centering{\
     \epsfig{angle=270,figure=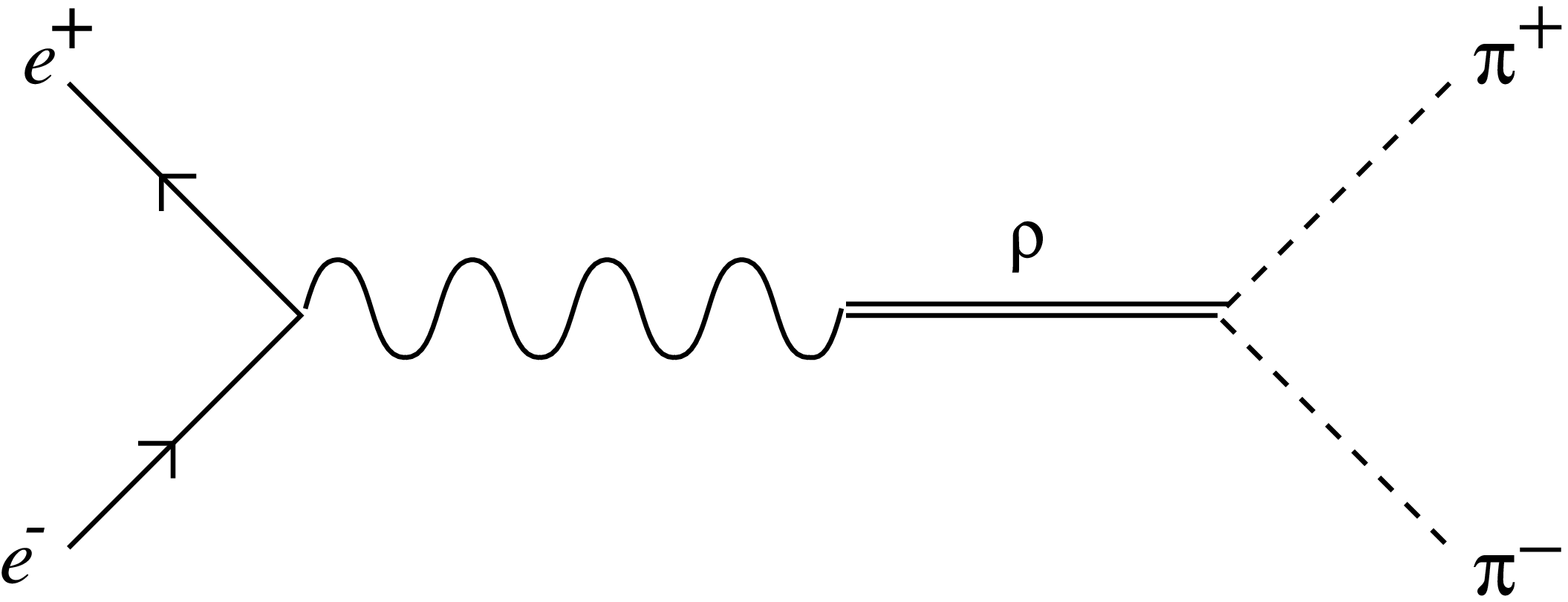,height=5cm}  }
\parbox{130mm}{\caption{VMD description of $e^+e^-\ra\pi^+\pi^-$.}
\label{vmdepem}}
\end{figure}

Early experiments measuring this cross-section
produced enough data around the $\rho$ resonance 
to enable the extraction of $|F_\pi(s)|^2$
and led to the development of the VMD model discussed earlier.
In the second representation of VMD the 
$e^+e^-\ra\pi^+\pi^-$ reaction is given by the process 
illustrated in Fig.~\ref{vmdepem}, which leads to the expression 
for the pion form-factor, as given in Eq.~(\ref{E2}),
\be
F_\pi(s)=\frac{-m_\rho^2}{s-m_\rho^2+im_\rho\Gamma_\rho}.
\label{interp2ff}
\ee
The reader will notice Eq.~(\ref{interp2ff}) differs slightly from
what would be naively expected from Eq.~(\ref{E2}) as the width
of the $\rho$-meson has been included. This will be fully discussed
later.

\subsection{The observation of \rw mixing}
\begin{figure}[htb]
  \centering{\
     \epsfig{angle=270,figure=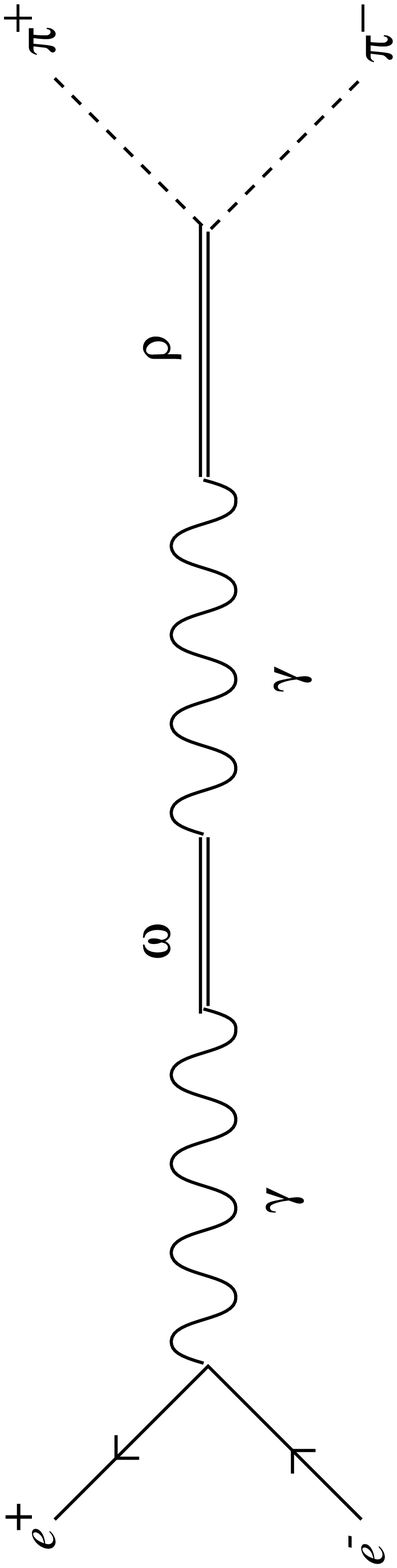,height=3cm}  }
\parbox{130mm}{\caption{Electromagnetic contribution to the 
$\omega$-resonance of $e^+e^-\ra\pi^+\pi^-$. }
\label{vmdem}}
\end{figure}

\mb{.5cm}
As more data was collected (for the reaction $e^+e^-\ra\pi^+\pi^-$ and other
related reactions such as $\pi^++p\ra\pi^+\pi^-+\Delta^{++}$)
and the resolution of the resonance curve improved, it became clear that there
was a kink
in the otherwise smooth curve observed around the mass of the
$\omega$-meson \cite{orsay1}. The strong interaction was not believed
to allow an
$\omega$ to decay to the pion pair, as to do so would violate G parity. 
Glashow suggested in 1961 \cite{Gla} that EM effects
mixed the two states of pure isospin,
$\rho_{I}$ and $\omega_{I}$, resulting in the mass 
eigenstates, $\rho$ and $\omega$, being
superpositions of the two initial fields.
The most obvious possibility, as this effect is only very small,
was via the process shown in Fig.~\ref{vmdem}.
He also commented that other EM mixing processes
such as $\rho_{I}\ra\gamma+\pi^0\ra\omega_{I}$ could not be ignored.
However, calculations revealed that the process shown in Fig.~\ref{vmdem}
is suppressed too much to account for what was seen in the experiment.
Being a second order electromagnetic effect it contributed 
only around 8 keV to the observed
partial width $\Gamma_{\omega\ra 2\pi}$=186 keV.

Hence it became necessary to abandon strict
conservation of  G-parity in the strong interaction. The explanation for the 
kink in the data was that the decay $\omega\ra 2\pi$ was {\em interfering}. 
It was even suggested \cite{Fub} that, as
the masses are so close, perhaps the $\rho$ and $\omega$ are just
decay modes (one to two pions and the other to three pions) of the one
particle, which, like the photon, did not possess a well-defined isospin.

However, a concerted effort to examine the decay $\omega\ra 2\pi$ 
concluded that there was not significant statistical evidence for the direct
decay \cite{LS}. It was suggested that perhaps, despite a possibly
substantial direct
decay rate, some process produced a cancellation giving a zero result. This 
argument for ignoring the direct decay was given a mathematical footing
\cite{R,GSR} that will be discussed in Sec.~ 3.4.

A way out of this problem seemed at hand with the strong symmetry breaking
theory of Coleman and Glashow \cite{CG}. This allowed for a mixing of the 
two mesons, introducing the quantity $\left<\right.\!\rho^0|M|\left.\omega
\right>$, where $M$ denotes the mass mixing operator which was 
taken to be a free parameter \cite{GFQ}. Ultimately, this mass-mixing has
its origins in the quark mass differences and EM effects, but 
there is as yet no definitive
derivation from QCD.

\begin{figure}[htb]
  \centering{\
     \epsfig{angle=270,figure=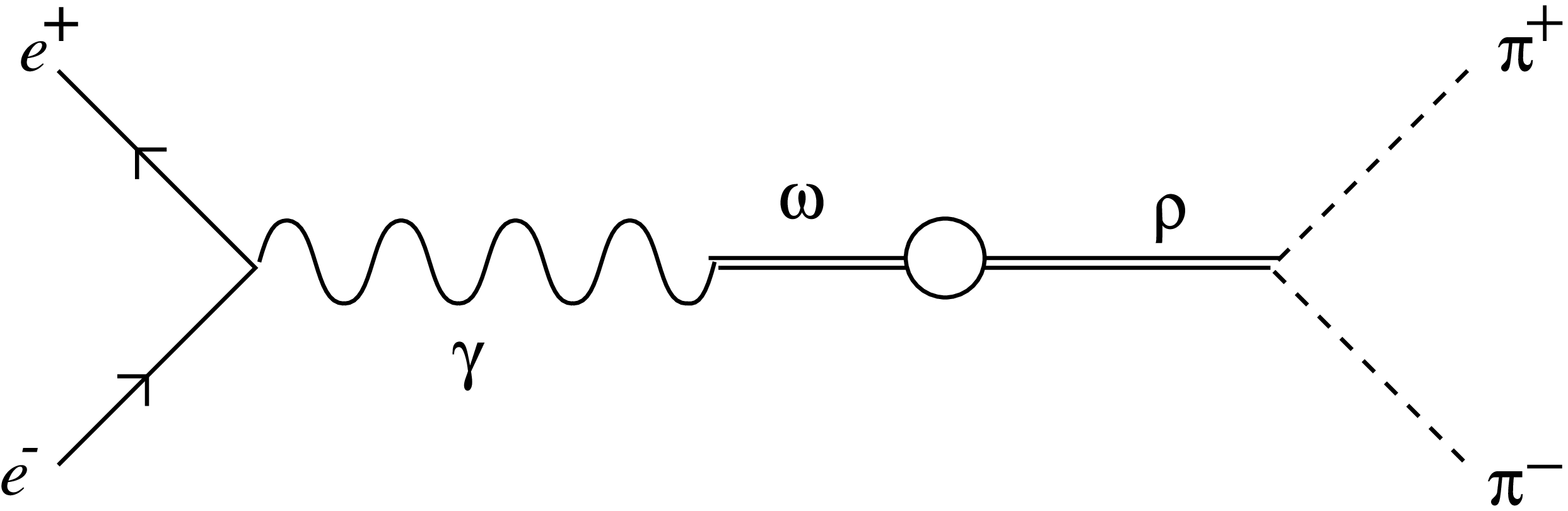,height=3cm}  }
\parbox{130mm}{\caption{\rw mixing contribution to  $e^+e^-\ra\pi^+\pi^-$.}
\label{rwmix}}
\end{figure}

\subsection{Quantum mechanical view of \rw mixing}
\label{quantum}
\mb{.5cm}
Our initial presentation of \rw mixing
will follow standard treatments \cite{R,SW}
originally due to Coleman and Schnitzer \cite{CS}. Although such methods are 
not usually employed today in the discussion of \rw mixing, they contributed
significantly to the development of the subject.

The vector meson propagator is given by
\be
D_{\mu\nu}(q^2)=\int d^4 x e^{-i q\cdot x}\left< 0|T\{V_\mu (x)
V_\nu (0)\}|0\right> 
\ee
which we can rewrite using the spectral representation \cite{Lu}
\be
D_{\mu\nu}(q^2)=\int_{s_0}^\infty dr \frac{\sigma(r)}{q^2-r}(g_{\mu\nu}-
\frac{q_\mu q_\nu}{r}) 
\label{spectral}
\ee
where $\sigma(r)$ is the spectral density of the vector states. From Eq.~(\ref{spectral})
we can define the propagator function, $D(q^2)$, such that \cite{R}
\be
D_{\mu\nu}(q^2)\equiv D(q^2)g_{\mu\nu}+\frac{1}{q^2}(D(0)-D(s))q_\mu q_\nu
\label{renardspectral}
\ee
where we define for convenience here $s\equiv q^2$.
We now write the propagator function in the following way
\be
D(s)=\frac{1}{s-W(s)}\,
\ee
where, in what follows, we shall regard $D$ and $W$ as operators.
The mass-squared operator, $W$, is a function of $s$ in general and we will
later use the form
\be
W(s)=m_0^2+\Pi(s)\;,
\label{renardW}
\ee
where $\Pi(s)$ is the self-energy operator with complex matrix elements and 
is related to the physical 
intermediate states (we shall discuss this in section \ref{rwzero})). 
The poles of the matrix elements of $D$ correspond to the physical
vector meson states. 

If we restrict our attention to the region near the $\omega$ mass, which
corresponds 
to a small energy range (of order $\Gamma_\omega$ or $m_\omega-m_\rho$), we can
safely neglect the $s$-dependence of $W$. The decay widths can thus
be taken as independent of $s$ in this region.

The physical states can be taken to be 
linear combinations of the pure isospin states, $|a_{I}\rangle$,
$a=\rho,\omega$, where  
\beas
 |\rho_{I}\rangle&\equiv& |1,0\rangle \\
 |\omega_{I}\rangle&\equiv& |0,0\rangle
\eeas
in the isospin basis, $|I,I_3 \rangle$. $W$ and $D$ would be diagonal
if there were no isospin-violating effects, but the existence 
of such effects produces
matrix elements which are not diagonal and the off-diagonal elements contain
the information about \rw mixing. Assuming time reversal invariance these
matrix elements are symmetric, though not real (and hence not necessarily
hermitian). The physical states are those which diagonalise $W$ and we
denote them by $|a\rangle$.  Either representation, the physical states
$|a\rangle$ or the isospin states $|a_{I}\rangle$ form a complete
orthonormal basis; i.e.,
\be
I = \sum_{a} |a\rangle \langle a| = \sum_{a_{I}} |a_{I}\rangle \langle a_{I}|
\ee
and
\be
\delta_{ab} = \langle a|b \rangle = \langle a_{I}|b_{I} \rangle .
\ee
Hence the two bases can be related by
\be
 |a\rangle = \sum_{b_{I}} |b_{I}\rangle \langle b_{I}|a\rangle
\label{eq:basetransform}
\ee
and
\be
 |a_{I}\rangle = \sum_{b} |b\rangle \langle b|a_{I}\rangle
\ee
We note here that we {\em define} the left eigenvectors $\langle a|$ by these
definitions.  We will see later that the transformation matrix with elements
$\langle b_{I}|a\rangle$ is not unitary and hence
$\langle a|\neq(|a\rangle)^{\dagger}$.
Naturally, $D(s)$ can be represented in either basis, for example in the
physical or mass basis
\be
  D = \sum_{a,b} |a\rangle\langle a| [s - W(s)]^{-1}
        |b\rangle\langle b|
\label{eq:Hisobasis}
\ee
with a similar expression using the  basis $|a_{I}\rangle$.  Since
the physical states are those that diagonalise $D$ and $W$ we can write
\be
  \langle a| W |b\rangle = \delta_{ab} z_{a}
\ee
and Eq.~(\ref{eq:Hisobasis}) becomes
\be
  D = \sum_{a} \frac{|a\rangle\langle a|}{s - z_a}
\label{hdef}
\ee

Since the mixing is observed to be small, we approximate the transformation
between the two bases given in Eq.~(\ref{eq:basetransform}) by
\bea
  |\rho\rangle &=& |\rho_{I}\rangle - \epsilon |\omega_{I}\rangle 
\label{eq:states1} \\
  |\omega\rangle &=& |\omega_{I}\rangle + \epsilon |\rho_{I}\rangle
\label{eq:states}
\eea
where $\epsilon$ is a small, complex mixing parameter.  Here and in the
following we always work to first order in $\epsilon$.
In matrix form, we write
\be
{\cal C} \equiv
  \left( \begin{array}{cc}
        \langle\rho_{I}|\rho\rangle   & \langle\omega_{I}|\rho\rangle \\
        \langle\rho_{I}|\omega\rangle & \langle\omega_{I}|\omega\rangle
         \end{array} \right)
= \left( \begin{array}{cc}
                1 & -\epsilon \\
        \epsilon  & 1
         \end{array} \right)
\ee
and
\be
{\cal W}_{I} \equiv
  \left( \begin{array}{cc}
        \langle\rho_{I}|W|\rho_{I}\rangle   & \langle\rho_{I}|W|\omega_{I}\rangle \\
        \langle\omega_{I}|W|\rho_{I}\rangle & \langle\omega_{I}|W|\omega_{I}\rangle
         \end{array} \right)
\ee
where the script letters are used to denote matrices.
The physical basis $|\rho\rangle$, $|\omega\rangle$ diagonalises $W$ so we
have
\be
{\cal W} = {\cal C} {\cal W}_{I} {\cal C}^{-1}
= \left( \begin{array}{cc}
        z_{\rho} & 0 \\
        0        & z_{\omega}
         \end{array} \right)
\ee
{}from which we deduce, neglecting all terms of order $\epsilon^2$ and
$\epsilon\langle\rho_{I}|W|\omega_{I}\rangle$ and observing that ${\cal W}_{I}$ must
be symmetric so that
$\langle\omega_{I}|W|\rho_{I}\rangle=\langle\rho_{I}|W|\omega_{I}\rangle$, 
\be
\epsilon=\frac{\langle\rho_{I}|W|\omega_{I}\rangle}{z_{\omega}-z_{\rho}}.
\ee
Since $z_a$ corresponds to the square of the complex mass, it is convenient
to write \cite{HS1}
\bea
\nonumber
z_a&=&(m_a- i \Gamma_a/2)^2 \\
&\simeq& m_a^2-i m_a \Gamma_a \,,
\label{evals}
\eea
where $\Gamma_a$ is the decay width of particle $a$ which was seen in the 
form-factor given in Eq.~(\ref{interp2ff}). Hence we have
\be
\epsilon=\frac{\langle\rho_{I}|W|\omega_{I}\rangle}
  {m^2_{\omega}-m^2_{\rho}-i(m_{\omega}
\Gamma_\omega-m_{\rho}\Gamma_\rho)}\,.
\label{epsilon}
\ee
Now we recall that $W$ is in general momentum dependent, and that
we neglected the momentum dependence as a simplification (as we
were concerned with only a small region around the $\omega$ mass). 
Hence, Eq.~(\ref{epsilon}) and therefore
\rw mixing will, in general, be momentum dependent. Interestingly,
$\epsilon$ is seen to have the form (neglecting the $\omega$ width)
of a $\rho$ propagator evaluated at $s=m_\omega^2$, which we can compare
with the discussion surrounding Eq.~(\ref{frozeq}).

The amplitude for any process involving intermediate vector states 
(which mix) will involve matrix elements of the vector meson propagator
function, $D$, and can now be written as, using Eq.~(\ref{hdef}),
\be
\langle f|D(s)|i\rangle=\sum_a\frac{\langle f|a\rangle \langle a|
i\rangle}{s-m_a^2+im_a\Gamma}\,.
\ee
For the case of \ep, using the more popular second representation of VMD
(Eq.~(\ref{newvmd})) we have
\bea
\nonumber
{\cal M}_{fi}&=&\langle 2\pi|D(s)|e^+e^-\rangle \\
&=&\frac{\langle 2\pi |\rho\rangle\langle \rho 
|e^+e^-\rangle }{ s-m_\rho^2+im_\rho\Gamma }
+\frac{ \langle2\pi|\omega\rangle \langle \omega|e^+e^-\rangle }{ s-
m_\omega^2+im_\omega\Gamma }\,.
\label{cont}
\eea
It is from this that we can determine the Orsay phase, $\phi$ 
\cite{orsay1,orsay2}, which is the relative phase of the $\omega$ and $\rho$ 
Breit-Wigner
amplitudes for $e^+e^-\ra2\pi$. 

Comparing with Eq.~(\ref{eq:m}) we can identify the pion form factor to be
\bea
\nonumber
F_{\pi}(s)&=&\frac{g_{\rho\pi\pi}g_{ \rho \gamma}}{s-m^2_{\rho}+i
m_{\rho}\Gamma_\rho}+\frac{g_{\omega\pi\pi}g_{ \omega \gamma}}{s-
m^2_{\omega}+im
_{\omega}\Gamma_\omega} \\
&\equiv&g_{\rho\pi\pi}g_{ \rho \gamma}
\left[\frac{1}{s-m^2_{\rho}+im_{\rho}\Gamma_\rho} \right.
\left.+\xi e^{i\phi}\frac{1}{s-m^2_{\omega}+im_{\omega}\Gamma_\omega}\right],
\label{orsayform}
\eea
where
\[
g_{\rho \gamma}=\frac{m_\rho^2}{g_\rho}.
\]
Hence the quantity $\xi e^{i\phi}\equiv (g_{ \omega \gamma}/ 
g_{ \rho
\gamma})(g_{\omega\pi\pi}/ g_{\rho\pi\pi})$ governs the shape of the 
interference
and hence of the cross-section around the $\omega$ mass. 

In the remainder of the paper, we make use of the following notational
conveniences, all valid only when terms of order $\epsilon^2$ and
$\epsilon\langle\rho_I|W|\omega_I\rangle$ can be neglected:
\bea
  W_{\rho\rho} &\equiv& \langle\rho_I|W|\rho_I\rangle
        = \langle\rho|W|\rho\rangle = z_{\rho} \nonumber \\
  W_{\omega\omega} &\equiv& \langle\omega_I|W|\omega_I\rangle
        = \langle\omega|W|\omega\rangle = z_{\omega} \nonumber \\
  W_{\rho\omega} &\equiv& \langle\rho_I|W|\omega_I\rangle.
\eea

\subsection{The contribution of direct omega decay}
\label{omegadecay}
\mb{.5cm}
As had been suggested \cite{LS} the existence of a direct decay of the 
pure isospin state, $\omega_{I}\ra 2\pi$ may have little effect on the decay 
of the real $\omega$. An argument was given for this \cite{R,GSR},
and most modern calculations do not include
the contribution of the direct
decay of the $\omega$. It is useful to outline these arguments 
and examine whether they still hold for recent examinations
of \rw mixing.

The coupling of the physical $\omega$ to the two pion state can be expressed 
as (from Eq.~(\ref{eq:states}))
\be
{\cal M}_{\omega\pi\pi}={\cal M}_{\omega_{I}\pi\pi}+\epsilon 
{\cal M}_{\rho_{I}\pi\pi},
\label{cont1}
\ee
where $\epsilon$ is given by Eq.~(\ref{epsilon}). 
Neglecting the small mass difference of the two
mesons and the decay width of the $\omega$ allows us to approximate
$\epsilon$, given in Eq.~(\ref{epsilon}), by
\be
\epsilon=-i\left(\frac{{\cal R}e\:W_{\rho\omega}
+i{\cal I}m\:W_{\rho\omega}}{m_\rho \Gamma_\rho}\right).
\label{epsy}
\ee

\begin{figure}[htb]
  \centering{\
     \epsfig{angle=270,figure=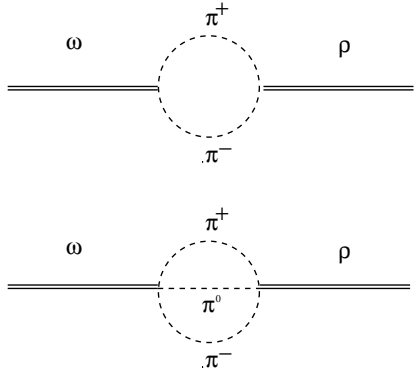,height=5cm}  }
\parbox{130mm}{\caption{Physical intermediate states contributing to 
\rw mixing.} 
\label{fig:int}}
\end{figure}

Now assuming the $\omega_I$ is able to couple to two pions (afterall,
some mechanism is required for CSV), i.e., 
${\cal M}_{\omega_{I}\pi\pi}\neq 0$, we
would have the mixing interaction, shown in Fig.~\ref{fig:int}, 
contributing to $W_{\rho\omega}$, and hence also to $\epsilon$.

\begin{figure}[htb]
  \centering{\
     \epsfig{angle=270,figure=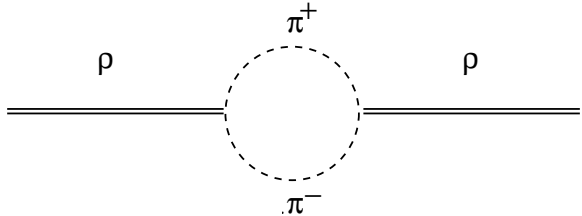,height=3cm}  }
\parbox{130mm}{\caption{Contribution of a pion loop to the $\rho$ 
self-energy.}
\label{fig:rho}}
\end{figure}

We can determine the contribution to $W_{\rho\omega}$ from
$\rho\ra\pi\pi\ra\omega$. To do this, however, it is first useful to consider
the analogous case for the simpler $\rho\pi$ system. The self energy of the 
$\rho$, $W_{\rho\rho}$, is generated by a virtual pion loop as 
in Fig.~\ref{fig:rho},
which modifies the $\rho$ propagator in the following way
\bea
\frac{1}{q^2-m_0^2}&\ra &\frac{1}{q^2-W_{\rho\rho}(q^2)} \\
	&\simeq& \frac{1}{q^2-m_\rho^2+im_\rho\Gamma_\rho},
\eea
where $m_0$ is the bare mass, $m_\rho$ the renormalised mass and
$\Gamma_\rho$ the width of the $\rho$-meson.
We now have the definition of the imaginary part of Fig.~\ref{fig:rho},
\be
{\cal I}m\;W_{\rho\rho}\equiv-m_\rho\Gamma_\rho(q^2),
\ee
where $\Gamma_\rho(m_\rho^2)$ is the decay width of the $\rho$.
Similarly we can determine the imaginary part of the one-loop diagram
shown in Fig.~\ref{fig:int}, which contributes to ${\cal I}m\;W_{\rho\omega}$.
If following the analysis of Renard \cite{R} we assume that the pure
isospin decay amplitudes are related by
\be
{\cal M}_{\rho\omega}=\frac{g_{\omega_I\pi\pi}}{g_{\rho_I\pi\pi}}
{\cal M}_{\rho\rho},
\ee
we have
\be
W_{\rho\omega}=\frac{g_{\omega_I\pi\pi}}{g_{\rho_I\pi\pi}}W_{\rho\rho},
\label{ratio}
\ee
and hence
\bea
\nonumber
{\cal I}m\;W_{\rho\omega}&=&\frac{g_{\omega_I\pi\pi}}{g_{\rho_I\pi\pi}}
{\cal I}m\;W_{\rho\rho} \\
&=&-\frac{g_{\omega_I\pi\pi}}{g_{\rho_I\pi\pi}}m_\rho\Gamma_\rho.
\label{imrho}
\eea
Substituting Eq.~(\ref{imrho}) into Eq.~(\ref{epsy}) and then 
substituting that into Eq.~(\ref{cont1}) we have
\be
g_{\omega\pi\pi}=g_{\omega_{I}\pi\pi}-i
\frac{{\cal R}e\:W_{\rho\omega}}{m_\rho \Gamma_\rho} g_{\rho_{I}\pi\pi}
+\frac{{\cal I}m\:W_{\rho\omega}}{m_\rho \Gamma_\rho} g_{\rho_{I}\pi\pi} 
\ee
Thus
\be
g_{\omega\pi\pi}=g_{\omega_{I}\pi\pi}-i
\frac{{\cal R}e\:W_{\rho\omega}}{m_\rho \Gamma_\rho} g_{\rho_{I}\pi\pi}
-g_{\omega_{I}\pi\pi}.
\ee
As can be seen the contribution from the decay of the $\omega_I$ is cancelled.

So, in summary, we allowed CSV through $\omega_I\ra\pi\pi$
(in the same form as $\rho_I\ra\pi\pi$), which
contributed to the mixing parameter, $\epsilon$, through the process
depicted in Fig.~\ref{fig:int}. We then found that the 
imaginary part of the single pion loop actually
cancelled the decay of the $\omega_I$ in the process $\omega\ra\pi\pi$.
Hence the decay of the the $\omega_I$ can be ignored.

However, the approximation of neglecting the $\omega$ 
width and the \rw mass
difference in Eq.~(\ref{epsy})
has recently been re-examined 
in detail \cite{MOW}. Without making
this approximation it 
has been found that the $g_{\omega_{I}\pi\pi}$ contribution survives
and can contribute significantly.

\subsection{Summary}
\mb{.5cm}

In this section we have concerned ourselves with the initial discovery
of the G-parity violating interactions of the $\omega$-meson which {\em could
not} be explained by electromagnetism alone. We also
reviewed the early theoretical attempts to
explain these processes. We described the development of
the notion of \rw mixing, a process which is still not entirely understood
at a fundamental level.

It is our purpose in the remainder of this report to develop a simple
framework for handling \rw mixing and show how to use it in practical 
calculations.

\section{Charge symmetry violation in nuclear physics}
\label{nucl}
\mb{.5cm}

Before proceeding to discuss \rw mixing in greater detail it is important
to briefly review its importance in nuclear physics.

There are a number of fine reviews of charge symmetry and the insight
which the small violations of it can give us concerning strongly interacting 
systems \cite{HM,b,Ger}. It would be inappropriate to go over that material at
length. Our objective here is simply to recall a few key examples 
where \rw mixing
is believed to play an important role. In this way we provide a framework 
within which our consideration of meson mixing and VMD may be viewed.

The charge symmetry breaking interaction of most interest in nuclear physics
has typically been the so-called class-III force \cite{HM} which
has the form,
\be
V^{\rm III}\equiv (\tau_{1z}+\tau_{2z})v_3(\ti{r},\ti{\sigma_1},
\ti{\sigma_2}).
\ee
This is responsible for the difference between the $nn$, (Coulomb corrected)
$pp$ and $np$ scattering lengths. It also contributes to a difference between 
the masses of mirror nuclei, the famous Okamoto-Nolen-Schiffer (ONS) 
anomaly \cite{ONS}. Given our ability to solve the three-body problem,
the $ ^3{\rm He}- ^3{\rm H}$ mass difference is the most precisely studied
example. After correcting for the EM interaction and the free $n-p$ mass
difference there remains some 70 keV to be explained in terms of a 
charge symmetry violating force \cite{WIS}. The class-III force associated
with \rw mixing predicts 90$\pm$14 keV \cite{WIS} which is in good
agreement.

For heavier nuclei the EM corrections are much more difficult to calculate
accurately. Nevertheless, after the best estimates have been made a CSV
mass difference remains which grows with nuclear mass number, $A$. 
As illustrated in Table~\ref{table} of the results of
Blunden and Iqbal \cite{BI} (taken from Ref.~\cite{MO})
a microscopic $NN$ potential, including CSV effects, can account for most
of this discrepancy --- at least for low $j$. 
Once again, \rw mixing appear to be
responsible for the majority (roughly 90\%) of the calculated effect.

\begin{table}[htb]
\begin{center}
\begin{tabular}{|cc|cc|cc|}
\hline\hline
\multicolumn{2}{|c|}{Nuclear Level} 
& \multicolumn{2}{c|}{Required CSV (keV)} &
 \multicolumn{2}{c|}{Calculated CSV (keV)} \\
&& DME &SkII   & total & $\rho^0-\omega$ \\
\hline
15& p$^{-1}_{3/2}$ & 250 & 190 & 210 & 182 \\
&p$^{-1}_{1/2}$ & 380 & 290 & 283 & 227 \\
\hline
17 & d$_{5/2}$ & 300 & 190 & 144 & 131 \\
& 1s$_{1/2}$ & 320 & 210 & 254 & 218 \\
& d$_{3/2}$ & 370 & 270 & 246 & 192 \\
\hline
39 & 1s$^{-1}_{1/2}$ & 370 & 270 & 337 & 290 \\
& d$^{-1}_{3/2}$ & 540 & 430 & 352 & 281 \\
\hline
41 & f$_{7/2}$ & 440 & 350 & 193 & 175 \\
& 1p$_{3/2}$ &380 &340 & 295 & 258 \\
& 1p$_{1/2}$ &410 &330 & 336 & 282 \\
\hline\hline
\end{tabular}
\parbox{130mm}{\caption{Summary of CSV in the single particle levels for
several light nuclei in comparison with the theoretical expectations
\protect\cite{BI} --- from \protect\cite{MO}.}
\label{table}}
\end{center}
\end{table}

There has also been considerable experimental activity in the past few years
\cite{Ab,Vig} concerning the class-IV force:
\be
V^{\rm IV}=(\tau_{1z}-\tau_{2z})(\ti{\sigma}_1-\ti{\sigma}_2)\cdot \ti{L}v(r)
+(\vec{\tau}_1\times\vec{\tau}_2)_z(\ti{\sigma}_1\times\ti{\sigma}_2)\cdot
\ti{L} w(r),
\ee
where vectors in isospin-space have been denoted by overhead arrows, and
those in position space by underlining.
Such a force only affects the $np$ system where it mixes the spin
singlet and triplet channels \cite{MilWil}. It turns out that at TRIUMF 
energies \cite{Ab} the measurement is insensitive to \rw mixing and
agrees well with the theoretical expectations \cite{MilWil}. On the other hand,
at the IUCF energy \cite{Vig} the data agrees well with the theoretical
prediction \cite{MilWil,Hol}, about half of which 
can be explained in terms of \rw mixing.
Unfortunately, the experimental error is such that this is only a 
$1.5$
standard deviation effect. It would be very informative to reduce the errors
by a factor of 2-3 in the IUCF energy region.

\begin{figure}[htb]
  \centering{\
     \epsfig{angle=270,figure=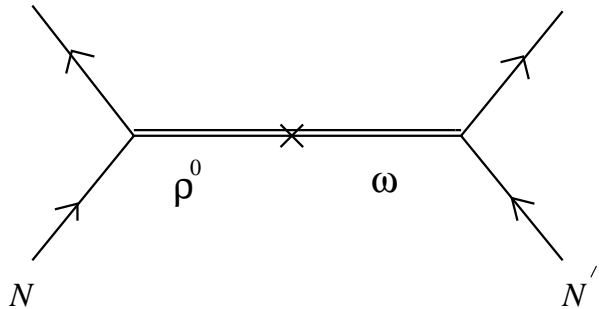,height=4cm}  }
\parbox{130mm}{\caption{CSV in the nuclear potential resulting
{}from \rw mixing.}
\label{rhoex}}
\end{figure}

Clearly there are a number of examples where CSV in nuclear physics seems
to require the contribution to the $NN$ force arising from \rw mixing.
In order to calculate such a force one must take the Fourier transform
of the Feynman diagram shown in Fig.~\ref{rhoex}.

Schematically this involves \cite{CSM}:
\be
V_{\rm CSV}(r)\propto\frac{1}{r}\int_{-\infty}^{\infty}dq\;q
\frac{\sin(qr)\Pi_{\rho\omega}(-q^2)}{(q^2+m_\rho^2)(q^2+m_\omega^2)}
\ee
where $\Pi_{\rho\omega}(-q^2)$ is the \rw mixing amplitude
{\em in the space-like region}. Traditionally this has been evaluated using
contour integration and keeping only the poles associated with the vector
meson propagator. That is, $V_{\rm CSV}$ is proportional to 
$\Pi_{\rho\omega}(m_\rho^2)$, the mixing amplitude at the $\rho$ (or
the $\omega$) pole. If $\Pi_{\rho\omega}$ were to vary rapidly between the 
time-like and space-like regions, as first suggested by Goldman
{\it et al.} \cite{GHT}, this would be a very bad approximation.
Indeed, if the behaviour found by Goldman {\it et al.} (discussed in the
next section) were correct, \rw mixing would contribute little or
nothing in the example we have just considered \cite{IN}. One would then be 
faced with the task of finding alternative, possibly quark-level
\cite{lb,la,SMG,MSG}, explanations. In any case, one would be forced to
re-examine the understanding of nuclear matter at a fairly fundamental
level.

\section{The behaviour of \rw mixing}
\mb{.5cm}
The various proposed mechanisms for \rw mixing 
(as, for example, the pion loops of Fig.~\ref{fig:int}) would have inescapably
led to the conclusion that it was a momentum dependent process. However no
direct calculations were ever made of these loop diagrams. 

In the early studies of \rw mixing
the mixing parameter $\epsilon$ (c.f. Eqs.~(\ref{eq:states1}) and 
(\ref{eq:states}))
was never precluded 
{}from being momentum dependent. Unfortunately,  experimental 
limitations meant there was little hope that much could
be known about $\epsilon$ away from the $\rho$ mass. Faced with this
constraint it seemed sensible to devote ones energies to finding out
as much as possible about the mixing process at $q^2=m_\omega^2$. The 
information came exclusively from the decay $\omega\ra\pi^+\pi^-$, i.e. two
pion production at the $\omega$ mass point, which (as we discussed)
was believed to be entirely due to mixing. (Note, though, that
recently there has been
some discussion of the experimentally more difficult $\rho\ra 3\pi$ decay
\cite{BC}.)
Renard \cite{R} gives a discussion of the behaviour of the mixing, in terms
of $W$ in Eq.~(\ref{renardW}). He explains that there were two approximations
made for the momentum dependence. The first was to ignore any momentum
dependence, the second was to assume it was linear (in which case it would
vanish for $s=0$, as predicted in section \ref{rwzero}).

As time went by any thought of $\epsilon$ being anything other than a fixed
parameter that could be cleanly extracted from processes
involving the two pion decay of the $\omega$ simply fell by the wayside (much
like the first representation of VMD).

While this has little effect
for something such as the EM form factor of the pion, its eventual application
\cite{CB}
to the spacelike world of nuclear physics 
where it has been incorporated into the meson
exchange model was cause for if not concern, at
least caution. However, the success of this assumption 
(outlined in Sec.~ \ref{nucl}) has been seen as a
compelling justification.

The question of momentum dependence in \rw mixing was first asked by Goldman,
Henderson and Thomas (GHT) \cite{GHT} and has generated a significant
amount of work. The initial GHT model was relatively simple. The vector mesons
were assumed to be quark-antiquark composites, and the mixing was generated
entirely by the small mass difference between the up and down quark masses.
The mesons coupled to the quark loop via a form-factor $F(k^2)$ where $k_\mu$
is the free momentum of the quark loop, which models the finite size of the
meson substructure. Free Dirac propagators were used for the quarks, thus
ignoring the question of confinement. More recent work \cite{KTW,MTRC}
has modelled confinement by using quark propagators which are entire
(i.e. they do not have a pole in the complex $q^2$ plane
and thus the quarks are never on mass-shell). 
The vector mesons couple to conserved currents which, as will be shown later,
leads to a node in the mixing when the momentum squared ($q^2$) of the
meson vanishes \cite{HOC}. 
A gauge invariant model, will produce a node at $q^2=0$ (see next section).
However the form-factors used in the GHT model spoil gauge invariance,
and thus their node is shifted slightly away from $q^2=0$.

The use of an intermediate nucleon loop \cite{PW} as the mechanism driving \rw
mixing (relying on the mass difference between the neutron and proton) avoids
the worries of quark confinement, as well as enabling one to use well-known
parameters in the calculation (masses, couplings, etc). This model has a node
for the mixing at $q^2=0$. Mitchell {\it et al.} \cite{MTRC} concluded that in
their bi-local theory (where the meson fields are composites of quark
operators, e.g. $\omega_\mu(x,y)\sim \overline{q}(y)i\gamma_\mu q(x)$) the
quark loop mechanism alone generates an insignificant charge symmetry breaking
potential and suggest a pion loop contribution should be examined
\cite{Mitchell}, which is interesting in the light of our discussion (Sec.~3.4)
about the contribution of the direct $\omega$ decay. Subsequent calculations
using the Nambu--Jona-Lasinio model \cite{RF}, chiral perturbation theory 
\cite{chiralp1,chiralp2,chiralp3}, QCD sum rules
\cite{chiralp2,HHMK,sum-rules1,sum-rules2} 
and quark models \cite{shakin} have explored
aspects of \rw mixing, including its momentum dependence.

Iqbal and Niskanen \cite{IN} have studied the effect of a varying \rw
mixing for neutron-proton scattering. Using a model for the variation 
\cite{HHMK} they conclude that it would significantly alter our
understanding of how to model the charge-symmetry breaking effects in the
strong nuclear interaction.

\subsection{General Considerations}
\label{rwzero}
\mb{.5cm}
We review our proof \cite{HOC} that the
mixing amplitude vanishes at $q^2=0$ in any effective Lagrangian
model (e.g., ${\cal L}(\vec\rho,\omega,\vec\pi,\bar\psi,\psi,\cdots)$),
where there are no
explicit mass mixing terms (e.g., $m^2_{\rho\omega}\rho^0_\mu\omega^\mu$ or
$\sigma\rho^0_\mu\omega^\mu$ with $\sigma$ some scalar field)
in the bare Lagrangian and where the 
vector mesons
have a local coupling to conserved currents which satisfy the usual vector
current commutation relations. 
The boson-exchange model of Ref.~\cite{PW}
where, e.g., $J^\mu_\omega = g_\omega\bar{N}\gamma^\mu N$, is one 
particular
example. It follows
that the mixing tensor (analogous to the full
self-energy
function for a single vector boson such as the $\rho$ \cite{BLP})
\be
C^{\mu\nu}(q)=i\int d^4x\,e^{iq\cdot x}\left<0\right|T(J^\mu_{\rho}(x)
J^\nu_\omega(0))\left|0\right>.
\label{eq:tensor}
\ee
is transverse. 
Here, the operator $J^\mu_\omega$ is the operator appearing in the
equation of motion for the field operator $\omega$ ---
c.f. Eq.~(\ref{eqmot}).  Note that when $J^\mu_\omega$ is a conserved current
then $\partial_\mu J^\mu_\omega=0$, which ensures that the Proca equation
leads to the same subsidiary condition as the free field case,
$\partial_\mu \omega^\mu=0$ (see, e.g., Lurie, pp.~186--190 \cite{Lu}, or
other field theory texts \cite{BjD,spec}).
The operator $J^\mu_\rho$ is similarly defined.
We see then that $C^{\mu\nu}$ can be written in the form,
\be
C^{\mu\nu}(q)=\left(g^{\mu\nu}-{q^\mu q^\nu \over q^2}\right)C(q^2)\,.
\label{transverse}
\ee

{From} this it follows that the one-particle-irreducible self-energy or
polarisation, $\Pi^{\mu\nu}(q)$ (defined through Eq.~(\ref{piC}) below),
must also be transverse \cite{BLP}.  
The essence of the argument below is that since 
there are no massless, strongly interacting vector particles $\Pi^{\mu\nu}$
cannot be singular at $q^2=0$ and   therefore
$\Pi(q^2)$ (see Eq.~(\ref{pi}) below)
must vanish at $q^2=0$, as suggested for the pure $\rho$ case
\cite{HFN}. As we have already noted this is something that was approximately
true in all models, but guaranteed only in Ref.~\cite{PW}.

Let us briefly recall the proof of the transversality of $C^{\mu\nu}(q)$.
As shown, for example, by Itzykson and Zuber (pp.~217--224) \cite{IZ},
provided we use covariant 
time-ordering the divergence of $C^{\mu\nu}$ leads to a naive commutator of
the appropriate currents
\bea
\nonumber
q_\mu C^{\mu\nu}(q)&=&-\int d^4x\,e^{iq\cdot x}\pa_\mu \;\{\theta(x^0)
\left<0\right|J^\mu_\rho(x)J^\nu_\omega(0)\left|0\right> \\
&\,\,&+\;\theta(-x^0)\left<0\right|J^\nu_\omega(0) J^\mu_\rho(x)
  \left|0\right>\} \\
&=& -\int d^3x\,e^{i\vec{q}\cdot \vec{x}}
\left<0\right|[J^0_\rho(0,\vec{x}),J^\nu_\omega(0)]\left|0\right>_{\rm naive}.
\label{commutator}
\eea
That is, there is a cancellation between the seagull and Schwinger terms.
Thus, for any model in which the isovector- and isoscalar- vector currents
satisfy the same commutation relations as QCD we find
\be
q_\mu C^{\mu\nu}(q)=0.
\ee
Thus, by Lorentz invariance, the tensor must be of the form
given in Eq.~(\ref{transverse}).

For simplicity we consider first the case of a single vector meson (e.g. a 
$\rho$ or $\omega$) without channel coupling. For such a system one can
readily see that since $C^{\mu\nu}$ is
transverse the one-particle irreducible self-energy, $\Pi^{\mu\nu}$, defined
through \cite{BLP}
\be
\Pi^{\mu\alpha}D_{\alpha\nu}=C^{\mu\alpha} D^0_{\alpha\nu}
\label{piC}
\ee
(where $D$ and $D^0$ are defined below) is also transverse.
 Hence
\be
\Pi^{\mu\nu}(q)=\left(g^{\mu\nu}-{q^\mu q^\nu \over q^2}\right)\Pi(q^2)\,.
\label{pi}
\ee

We are now in a position to establish the behaviour of the
scalar function, $\Pi(q^2)$. In
a general theory of massive vector bosons coupled to a conserved current, the
bare propagator has the form (compared to Eq.~(\ref{photonprop}) 
for the photon)
\be
D^0_{\mu\nu}=\left(-g_{\mu\nu}+{q_\mu q_\nu \over m^2}\right){1 \over
{q^2-m^2}}\ee
whence
\be
(D^0)^{-1}_{\mu\nu}=(m^2-q^2)g_{\mu\nu}+q_\mu q_\nu.
\ee
The polarisation is incorporated in the standard way to give the dressed
propagator
\be
iD_{\mu\nu}=iD_{\mu\nu}^0+iD_{\mu\alpha}^0i\Pi^{\alpha\beta}iD_{\beta\nu}^0
+\cdots
\label{propexpan}
\ee
We now use the operator identity of Eq.~(\ref{opidentity}) to give
\bea
\nonumber
D^{-1}_{\mu\nu}&=&(D^0)^{-1}_{\mu\nu}+\Pi_{\mu\nu} \\
&=&(m^2-q^2+\Pi(q^2))g_{\mu\nu}+\left(1-{\Pi(q^2)\over q^2}\right)q_\mu q_\nu.
\label{eq:inverse}
\eea
Thus the full propagator has the form
\be
D_{\mu\nu}(q)={-g_{\mu\nu}+\left(1-\Pi(q^2)/q^2\right)(q_\mu q_\nu/
m^2)\over q^2-m^2-\Pi(q^2)}.
\label{prop1}
\ee

Having established this form for the propagator, we wish to compare it with
the Renard spectral representation of the propagator given by 
Eq.~(\ref{renardspectral}).
By comparing the coefficients of $g_{\mu\nu}$ in Eqs.~(\ref{prop1}) and
(\ref{renardspectral}) we deduce
\be
D(q^2)={-1\over {q^2-m^2-\Pi(q^2)}}\,,
\ee
while from the coefficients of $q_\mu q_\nu$ we have
\bea
\nonumber
{\left(1-{\Pi(q^2)/ q^2}\right)\over {(q^2-m^2-\Pi(q^2))m^2}}&=&{1\over
q^2}(D(0)-D(q^2)) \\
&=& {1\over q^2}{q^2 +\Pi(0)-\Pi(q^2) \over (m^2+\Pi(0))(q^2-m^2-\Pi(q^2))},
\eea
{}from which we obtain
\be
\frac{\Pi(0)}{q^2}(q^2-m^2-\Pi(q^2))=0 \,\,\,\,\,\,\,\, ,\,\,\,\,\,\forall q^2
\ee
and thus
\be
\Pi(0)=0.
\ee
This is an important constraint on the self-energy function, namely that
$\Pi(q^2)$ should vanish as $q^2\ra 0$ at least as fast as $q^2$.

While the preceding discussion dealt with the single channel case, for
$\rho-\omega$ mixing we are concerned with two coupled channels. Our
calculations therefore involve matrices. As we now demonstrate, this does not
change our conclusion.

The matrix analogue of Eq.~(\ref{eq:inverse}) is
\be
D^{-1}_{\mu\nu}=\left(\begin{array}{cc}
m_\rho^2g_{\mu\nu}+(\Pi_{\rho\rho}(q^2)-q^2)T_{\mu\nu} &
\Pi_{\rho\omega}(q^2)T_{\mu\nu} \\   
\Pi_{\rho\omega}(q^2)T_{\mu\nu} & m_\omega^2g_{\mu\nu}+
(\Pi_{\omega\omega}(q^2)-q^2)T_{\mu\nu}
\end{array} \right),
\ee
where we have defined $T_{\mu\nu}\equiv g_{\mu\nu}-(q_\mu q_\nu/q^2$)
for brevity.
By obtaining the inverse of this we have the two-channel propagator
\be
D_{\mu\nu}=\frac{1}{\alpha}\left(\begin{array}{cc}
s_\omega g_{\mu\nu}+a(\rho,\omega) q_\mu q_\nu &  \Pi_{\rho\omega}(q^2)
T_{\mu\nu} \\
 \Pi_{\rho\omega}(q^2)T_{\mu\nu} & s_\rho g_{\mu\nu}+a(\omega,\rho) q_\mu q_\nu
\end{array} \right),
\label{prop3}
\ee
where 
\bea
s_\omega&\equiv&q^2-\Pi_{\omega\omega}(q^2)-m_\omega^2 \\
s_\rho &\equiv&q^2-\Pi_{\rho\rho}(q^2)-m_\rho^2 \\
a(\rho,\omega)&\equiv&\frac{1}{q^2 m_\rho^2}\{\Pi_{\rho\omega}^2(q^2)-
[q^2-\Pi_{\rho\rho}(q^2)]s_\omega\} \\
\alpha &\equiv& \Pi_{\rho\omega}^2(q^2)-s_\rho s_\omega .
\label{alpha}
\eea
In the uncoupled case [$\Pi_{\rho\omega}(q^2)=0$] Eq.~(\ref{prop3}) clearly
reverts to the appropriate form of the one particle propagator,
Eq.~(\ref{prop1}), as desired.

We can now make the comparison between Eq.~(\ref{prop3}) and the Renard form
\cite{R} of the propagator, as given by Eq.~(\ref{renardspectral}). The transversality
of the off-diagonal terms of the propagator, demands that
$\Pi_{\rho\omega}(0)=0$. A similar analysis leads one to conclude the same for
$\Pi_{\rho\rho}(q^2)$ and $\Pi_{\omega\omega}(q^2)$.
Note that the physical $\rho^0$ and $\omega$ masses which arise from locating
the poles in the diagonalised propagator matrix $D^{\mu\nu}$ no longer 
correspond to exact isospin eigenstates (as in the discussion
of the historical treatment of \rw mixing, Sec.~\ref{quantum}).
To lowest order in CSV the 
physical $\rho$-mass is given by $m^{\rm phys}_\rho
=[m^2_\rho+\Pi_{\rho\rho}((m^{\rm phys}_\rho)^2)]^{1/2}$, i.e., the pole
in $D^{\mu\nu}_{\rho\rho}$.
The physical $\omega$-mass is similarly defined.

In conclusion, it is important to review what has and has not been 
established. There is no unique way to derive an effective field theory 
including vector mesons from QCD. Our result that $\Pi_{\rho\omega}(0)$ (as
well as $\Pi_{\rho\rho}(0)$ and $\Pi_{\omega\omega}(0)$) should vanish
applies to those effective theories in which: (i) the vector mesons have local
couplings to conserved currents which satisfy the same commutation relations
as QCD [i.e., Eq.~(\ref{commutator}) is zero]
and (ii) there is no explicit mass-mixing term in the bare
Lagrangian. This includes a broad range of commonly used, phenomenological
theories.
It does not include the model treatment of Ref.~\cite{MTRC} for example,
where the mesons are bi-local objects in a truncated
effective action.  However, it is interesting to note
that a node near $q^2=0$ was found in this model in any case.
The presence of an explicit mass-mixing term in the
bare Lagrangian will shift the
mixing amplitude by a constant (i.e., by $m_{\rho\omega}^2$).  We
believe that such a term will lead to difficulties in matching
the effective model  
onto the known behaviour of QCD in the high-momentum limit.

Finally the fact that $\Pi(q^2)$ is momentum-dependent or vanishes everywhere
in this class of models implies that the conventional {\it assumption}
of a non-zero, constant $\rho-\omega$ mixing amplitude remains
questionable.  This study then lends support to 
those earlier calculations, which we briefly discussed, where it was
concluded that the mixing may play a minor role in the explanation of CSV
in nuclear physics. It remains an
interesting challenge to find possible alternate mechanisms to describe
charge-symmetry violation in the $NN$-interaction
\cite{lb,la,alt1,alt2,alt3,alt4,alt5}.

\subsection{The mixed propagator approach to \rw mixing}
\label{mixprop}
\mb{.5cm}
Different authors parameterise the \rw mixing contribution to the pion 
form-factor in one of two ways. Using the matrix method we shall show here
the connection between these two models, both of which are first order
in charge symmetry breaking.

\begin{figure}[htb]
  \centering{\
     \epsfig{angle=270,figure=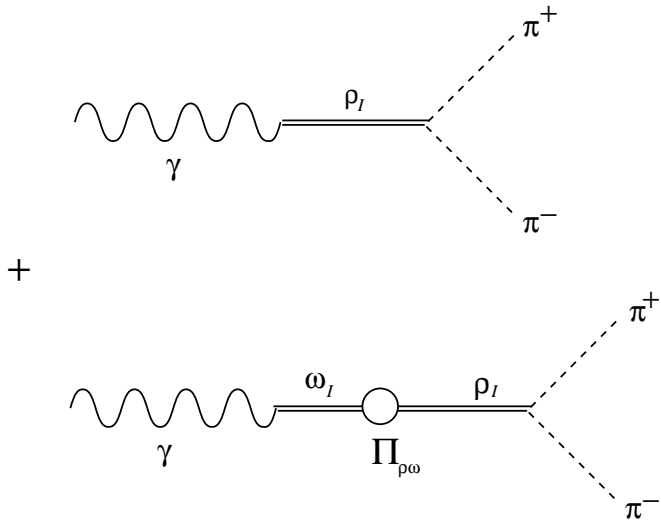,height=7cm}  }
\caption{The contribution of \rw mixing to the pion form-factor.}
\label{tony}
\end{figure}

Using a matrix notation, the Feynman amplitude for the process
$\gamma\ra\pi\pi$, proceeding via vector mesons, can be written in
the form
\begin{equation}
i\M_{\mu;\gamma\ra\pi\pi} =
  \left( \begin{array}{cc}
          i\M^{\nu}_{\rho_I\ra\pi\pi} & i\M^{\nu}_{\omega_I\ra\pi\pi}
         \end{array} \right)
  iD_{\nu\mu}
  \left( \begin{array}{c}
          i\M_{\gamma\ra\rho_I} \\
          i\M_{\gamma\ra\omega_I} 
         \end{array} \right)
\label{Fpimatrix}
\end{equation}
where the matrix $D_{\nu\mu}$ is given by Eq.~(\ref{prop3}) and the other
Feynman amplitudes are derived from either the VMD1 or VMD2 Lagrangian
(Eqs.~(\ref{vmdlag}) and (\ref{newvmd}).  Since we always couple the vector
mesons to conserved currents, the terms proportional to $q_{\mu}q_{\nu}$ in
the propagator (Eq.~(\ref{prop3})) can always be neglected.  If we assume that
the pure isospin state $\omega_I$ does not couple to two pions
($\M^{\nu}_{\omega_I\ra\pi\pi}=0$) then to lowest order in the mixing,
Eq.~(\ref{Fpimatrix}) is just
\begin{equation}
\M^{\mu}_{\gamma\ra\pi\pi} =
  \left( \begin{array}{cc}
          \M^{\mu}_{\rho_I\ra\pi\pi} & 0
         \end{array} \right)
  \left( \begin{array}{cc}
          1/s_{\rho} & \Pi_{\rho\omega}/s_{\rho}s_{\omega} \\
          \Pi_{\rho\omega}/s_{\rho}s_{\omega} & 1/s_{\omega}
         \end{array} \right)
  \left( \begin{array}{c}
          \M_{\gamma\ra\rho_I} \\
          \M_{\gamma\ra\omega_I} 
         \end{array} \right)
\label{ffampl}
\end{equation}
Expanding this just gives
\begin{equation}
  \M^{\mu}_{\gamma\ra\pi\pi} =
    \M^{\mu}_{\rho_I\ra\pi\pi} \frac{1}{s_{\rho}} \M_{\gamma\ra\rho_I}
    + \M^{\mu}_{\rho_I\ra\pi\pi} \frac{1}{s_{\rho}} \Pi_{\rho\omega}
        \frac{1}{s_{\omega}} \M_{\gamma\ra\omega_I}
\label{nondiag}
\end{equation}
which we recognise as the sum of the two diagrams shown in Fig.~\ref{tony}.

The couplings that enter this expression, through
$\M^{\mu}_{\rho_I\ra\pi\pi}$, $\M_{\gamma\ra\rho_I}$ and
$\M_{\gamma\ra\omega_I}$, always involve the unphysical pure isospin states
$\rho_I$ and $\omega_I$.  However, we can re-express Eq.~(\ref{nondiag}) in
terms of the physical states by first diagonalising the vector meson
propagator.  Following the same procedure as in Sec.~\ref{quantum}, we
introduce a diagonalising matrix
\begin{equation}
C = \left(\begin{array}{cc} 1 & \epsilon \\
-\epsilon & 1 \end{array} \right)
\end{equation}
where, to lowest order in the mixing,
\begin{equation}
\epsilon=\frac{\Pi_{\rho\omega}}{s_\rho-s_\omega}.
\label{epseq}
\end{equation}
We now insert identities into Eq.~(\ref{ffampl}) and obtain
\begin{eqnarray}
\nonumber
\M^{\mu}_{\gamma\ra\pi\pi} &=&
  \left( \begin{array}{cc}
          \M^{\mu}_{\rho_I\ra\pi\pi} & 0
         \end{array} \right)
  C C^{-1}
  \left( \begin{array}{cc}
          1/s_{\rho} & \Pi_{\rho\omega}/s_{\rho}s_{\omega} \\
          \Pi_{\rho\omega}/s_{\rho}s_{\omega} & 1/s_{\omega}
         \end{array} \right)
  C C^{-1}
  \left( \begin{array}{c}
          \M_{\gamma\ra\rho_I} \\
          \M_{\gamma\ra\omega_I} 
         \end{array} \right) \\
 &=&
  \left( \begin{array}{cc}
          \M^{\mu}_{\rho\ra\pi\pi} & \M^{\mu}_{\omega\ra\pi\pi}
         \end{array} \right)
  \left( \begin{array}{cc}
          1/s_{\rho} & 0 \\
          0 & 1/s_{\omega}
         \end{array} \right)
  \left( \begin{array}{c}
          \M_{\gamma\ra\rho} \\
          \M_{\gamma\ra\omega} 
         \end{array} \right)
\label{diagff}
\end{eqnarray}
where we have identified the physical amplitudes as
\begin{eqnarray}
\M^{\mu}_{\rho\ra\pi\pi}&=&\M^{\mu}_{\rho_I\ra\pi\pi}, \\
\M^{\mu}_{\omega\ra\pi\pi}&=&\epsilon \M^{\mu}_{\rho_I\ra\pi\pi}, \\
\M_{\gamma\ra\rho}&=&\M_{\gamma\ra\rho_I} - \epsilon \M_{\gamma\ra\omega_I}, \\
\M_{\gamma\ra\omega}&=&\M_{\gamma\ra\omega_I} + \epsilon \M_{\gamma\ra\rho_I}.
\end{eqnarray}
Expanding Eq.~(\ref{diagff}), we find
\begin{eqnarray}
\nonumber
  \M^{\mu}_{\gamma\ra\pi\pi} &=&
    \M^{\mu}_{\rho\ra\pi\pi} \frac{1}{s_{\rho}} \M_{\gamma\ra\rho}
   + \M^{\mu}_{\omega\ra\pi\pi} \frac{1}{s_{\omega}} \M_{\gamma\ra\omega} \\
 &=&
  \M^{\mu}_{\rho\ra\pi\pi} \frac{1}{s_{\rho}} \M_{\gamma\ra\rho}
   + \M^{\mu}_{\rho\ra\pi\pi} \frac{\Pi_{\rho\omega}}{s_\rho-s_\omega}
        \frac{1}{s_{\omega}} \M_{\gamma\ra\omega} ,
\label{diagff2}
\end{eqnarray}
which is the usually seen in older works. At first glance there seems to be a
slight discrepancy between Eqs.~(\ref{nondiag}) and (\ref{diagff2}). The
source of this is the definition used for the coupling of the vector meson to
the photon. The first, Eq.~(\ref{nondiag}), uses couplings to pure isospin
states, the second, Eq.~(\ref{diagff2}) uses ``physical" couplings (i.e.,
couplings to the mass eigenstates) which introduce a leptonic contribution to
the Orsay phase, as discussed by Coon {\it et al.}  \cite{CSM}. This phase is,
however, rather small.  If we assume
$\M_{\gamma\ra\rho_I}=3\M_{\gamma\ra\omega_I}$ and define the leptonic phase
$\theta$ by
\begin{equation}
\frac{\M_{\gamma\ra\omega}}{\M_{\gamma\ra\rho}}=\frac{1}{3}e^{i\theta}
\end{equation}
then, to order $\epsilon$,
\begin{equation}
\tan\theta=\frac{10\Pi_{\rho\omega}}{3m_\rho\Gamma_\rho} .
\end{equation}
This gives $\theta=5.7^o$ for $\Pi_{\rho\omega}=-4520$, as obtained by Coon
{\it et al.}  \cite{CSM}.  This small leptonic contribution to the Orsay phase
is the principal manifestation of diagonalising the \rw propagator.

\section{Phenomenological analysis of $F_\pi$}
\mb{.5cm}
In this section we discuss various methods for both fitting the pion
form factor and obtaining the numerical value of $\Pi_{\rho\omega}$. We
extract $\Pi_{\rho\omega}$ with a fit
to the pion form-factor (using VMD1), but, 
as will be seen, this is not the method
used to obtain the most widely quoted value.

Recent analysis of the \ep data give us an insight into
how successful the second formulation has been in describing the process.
We find that in both cases studied a non-resonant contribution has been
included to optimise the fit, in direct contrast with the spirit
of the second formulation. Following this we present an example of the 
use of the first formulation
to plot the curve for the cross-section of \ep. 
\newpage

\subsection{Recent fits}
\mb{.5cm}

Benayoun {\it et al.} \cite{BF} examine \ep in an effort to better understand
the process $\eta'\ra\pi^+\pi^-\gamma$, which requires a thorough
understanding of $\rho$ physics, i.e. how to effectively parameterise
it, and whether to include any non-resonant contributions to the
process. They are concerned primarily with fitting the data, relying on as
much experimental input as possible, rather than trying to test the behaviour
of a particular model for the process (which is our intention).

Their expression for the amplitude takes the form described in Eq.~(\ref{eq:m}),
with
\bea
\nonumber
F_\pi(s)(p_3-p_4)_\mu&=&\left[A(q^2)\Sigma_j \epsilon^j_\mu(\gamma)
\epsilon^{j*}_\nu(\gamma)
+\!g_{\rho\gamma}\frac{1}{m_\rho^2-q^2-im_\rho\Gamma_\rho(q^2)}
G_\rho
\Sigma_j \epsilon^j_\mu(\rho)\epsilon^{j*}_\nu(\rho) \right. \\
&+&\left.g_{\omega\gamma}
\frac{e^{i\phi}}{m_\omega^2-q^2-im_\omega\Gamma_\omega}G_\omega
\Sigma_j \epsilon^j_\mu(\omega)\epsilon^{j*}_\nu(\omega)\right](p_3-p_4)^\nu,
\label{omegabit}
\eea
where $j$ is the index associated with the helicity of the
polarisation vectors $\epsilon_\mu$. The first term,
$A(q^2)$, introduces their proposed
non-resonant contribution to $F_\pi(q^2)$.
Note the use of the momentum dependent width for the $\rho$ but not the 
$\omega$ (as its major decay channel, $3\pi$, is not included in the 
$2\pi$ data analysed, and one can, as an approximation,
ignore the momentum dependence of the width). The width was taken to 
be \cite{BF}
\be
\Gamma_\rho(q^2)=\Gamma_\rho \left(\frac{k(q^2)}{k(m_\rho^2)}\right)^3
\left(\frac{m_\rho}{\sqrt{q^2}}\right)^\lambda 
\label{width}
\ee
where $\lambda$ is a parameter for fitting,
\be
k(q^2)=\h\sqrt{q^2-4m_\pi^2}, \label{kof}
\ee
and
\be
G_V=\frac{1}{k^{3/2}(q^2)}\sqrt{6\pi m_V\sqrt{q^2} \Gamma_{V\pi^+\pi^-}}.
\label{coupling}
\ee
Note that, because of $k(q^2)$ in Eq.~(\ref{kof}), the width, 
$\Gamma_\rho(q^2)$
given in Eq.~(\ref{width}), will become imaginary below threshold, i.e.,
$q^2=4m_\pi^2$. Considering Eq.~(\ref{omegabit}), the width contribution
to the denominator of the propagator
(Eq.~(\ref{width})) will actually become real
and add to the mass term below threshold.
The use of a term such as $\theta(q^2-4m_\pi^2)$ in Eq.~(\ref{width})
would spoil the analytic continuation of
the propagator below threshold.
The width of the $\rho$ is almost entirely due to the two pion decay,
and thus the full width can be used in Eq.~(\ref{coupling}) to determine
$G_\rho$.
However
this is not the case for the $\omega$, so one has to make the appropriate
modification
\be
\Gamma_\omega(q^2)={\rm BR}(\omega\ra\pi^+\pi^-)\Gamma_\omega 
\left(\frac{k(q^2)}{k(m_\omega^2)}\right)^3
\left(\frac{m_\omega}{\sqrt{q^2}}\right)^\lambda,
\label{branch}
\ee
where ${\rm BR}(\omega\ra\pi^+\pi^-)$ is the branching ratio for the decay.
Note that Eq.~(\ref{omegabit}) uses the full width 
of the $\omega$, rather than the branching fraction as in Eq.~(\ref{coupling}).
This is because the width appearing in the propagator measures
the flux loss due to the decay of the particle irrespective of its decay 
channel. Conversely, $G_\omega$ describes the coupling of the $\omega$
to two pions only, so the partial width must be used.

The form-factor is thus
\be
F_\pi(q^2)=
A(q^2)-\frac{g_{\rho\gamma}}{s_\rho}G_\rho
-\frac{g_{\omega\gamma}}{s_\omega}e^{i\phi}G_\omega.
\label{beneq}
\ee
The resemblance of Eq.~(\ref{beneq}) to the older form given in 
Eq.~(\ref{orsayform}) is immediately apparent (i.e. the sum of
the $\rho$ contribution and an Orsay-phased $\omega$ contribution).
This can now be used in Eq.~(\ref{xsection}) to compute the cross-section.

Benayoun {\it et al.} \cite{BF} now proceed along two paths, 
using the accepted figures for
the $\omega$ as well as both leptonic decay widths (which are assumed
to be fairly well understood):

a) fitting the $\rho$ parameters and the Orsay phase, $\phi$, assuming
$A=0$;

b) fitting $A$ and leaving the $\rho$ mass fixed at the world average,
768.7$\pm$0.7 MeV, as it is believed to be less sensitive than the
width to parametrisation.

For the first case they arrive at values for the mass and width slightly
higher than usually found using the Gounaris-Sakurai model \cite{GS}
by, for example, Barkhov {\it et al.} \cite{Bark}.

For the second case, $A$ is assumed to take the form
\be
A(q^2)=-(c_0+c_2q^2+c_4q^4+\cdots).
\ee
The expansion is stopped as soon as the effect of the next term is negligible.
At this point Benayoun {\it et al.} pause to relect on the condition
$F_\pi(0)=1$. $A(0)$, as determined by their fit, would contribute 0.607
to $F_\pi(0)$. They dismiss the relevance of this as they are using ``an
expansion valid in the range $(\sqrt{s}=)\;[2m_\pi,m_{\eta'}]$.'' 
They go on to point
out that a good fit (which, in addition, reproduces values for the
parameters closer to the usual ones) can be obtained using
\be
A(q^2)=\frac{-1}{1+q^2/m_\rho^2}
\ee
which, of course, would contribute 1 to $F_\pi(0)$. They attempt to get
around this problem by commenting how it shows that values obtained
using extrapolation cannot be trusted. This serves to highlight the confusion
that surrounds the second representation of VMD away from mass-shell, and
more specifically, at $q^2=0$.

They conclude their investigation by saying that either
the $\rho$ mass is nearly degenerate with the $\omega$, or evidence
strongly suggests a non-VMD contribution to $F_\pi(q^2)$. Interestingly, they
say that the latter is suggested by the work of Bando {\it et al.} \cite{Ban}.

Bernicha, L\'{o}pez Castro and Pestieau (BCP) \cite{BCP}
obtain, conversely, significantly lower values than the world
average for $m_\rho$ and $\Gamma_\rho$. Their aim is to determine 
these two quantities in as model-independent a way
as possible. Their concern is that the values given by the Particle Data
Group (a slightly more recent list than referred to by Benayoun
{\it et al.}), 768.1 $\pm$ 0.5 and 151 $\pm$ 1.2 MeV are obtained
{}from different sources. The mass is obtained from photo-production and
$\pi^+ N\ra \rho N$, and the width from \ep. Thus it is possible
that there is some inconsistency due to different Breit-Wigner 
parametrisations. To rectify this they attempt to derive both from 
the available data of the cross-section for \ep \cite{Bark}. 
Using Eq.~(\ref{xsection}) they then plot the form-factor.

They assume the pion form-factor can be expressed in the form
\be
F_\pi(s)=\frac{A}{s-s_\rho}+B(s)
\label{first}
\ee
where $s_\rho$ is the position of the pole, $A$ the residue of the pole, and
$B(s)$ the non-resonant background near $s_\rho\equiv m_\rho^2-im_\rho 
\Gamma_\rho$. To include the contribution of the $\omega$ meson
they modify Eq.~(\ref{first}), in two ways:
\be
F_\pi(s)=\left(\frac{A}{s-s_\rho}+B(s)\right)
\left(1+y\frac{m_\omega^2}{s-s_\omega}\right)
\label{either} 
\ee
or
\be
F_\pi(s)=\frac{A}{s-s_\rho}\left(1+y\frac{m_\omega^2}{s-s_\omega}\right)
+B(s).
\label{or}
\ee
where $A$ is taken to be a constant, $A=-am_\rho^2$. With $B(s)=0$ these
equations reduce to the usual form, used, for example, by Barkhov
{\it et al.} \cite{Bark}.

Initially $B$ too is set to a constant, $b$, and the curve is fitted with
five parameters. Both parameterisations (Eqs.~(\ref{either})
and (\ref{or})) lead to essentially the same set of values for the parameters
that optimise the fit, so it is concluded that the \rw mixing and
background terms are only very weakly coupled. 

Interestingly, they fit the space-like data (obtained from $e\pi$ scattering
\cite{D}) using a form-factor given by
\be
F_\pi(s)=-\frac{am_\rho^2}{s-m_\rho^2}
\left[1+b\left(\frac{s-m_\rho^2}{m_\rho^2}\right)\right]^{-1},
\ee
which contains no contribution from the $\omega$. They say that it is 
negligible for $s<4m_\pi^2$. Whether this is because \rw mixing itself is much
smaller in this region, or merely because the $\omega$ pole does not
appear in this region is uncertain.

The calculations are then redone imposing $F_\pi(0)=1$. There is little
difference to the results, as would be expected; if $F_\pi(0)=1$
is a necessary condition, then any good fit should at least
come very close to fulfilling it.

They then examine the \rw mixing contribution more closely and consider the 
factor $-aym_\rho^2/(s-m_\rho^2+im_\rho\Gamma_\rho)$ being ``frozen'' at
a particular value of $s$, $\bar{s}$ say. This reproduces the type
of form-factor we encountered in Eq.~(\ref{orsayform})
and more recently in Eq.~(\ref{beneq}), which looks
simply like the sum of two Breit-Wigner amplitudes (one from the $\rho$
the other from the $\omega$) attenuated by the Orsay phase, $\phi$. 
This results in
\be
F_\pi(s)=-\frac{am_\rho^2}{s-m_\rho^2+im_\rho\Gamma_\rho}+
\delta(\bar{s}) e^{i\phi}\frac{m_\omega^2}{s-m_\omega^2+im_\omega\Gamma_\omega}
+b,
\label{frozeq}
\ee
where
\be
\delta (\bar{s}) e^{i\phi(\bar{s})}=
-\frac{am_\rho^2}{\bar{s}-m_\rho^2+im_\rho\Gamma_\rho}.
\label{expon}
\ee
There is little theoretical reason to do this, but it does explain the origin
of the Orsay phase, $\phi$, and is a reasonable approximation
(as the mixing is only noticeable around resonance). The reader familiar
with Sec.~\ref{mixprop} can recall the relationship between the
two formulations of the mixing, as outlined in Eqs.~(\ref{nondiag}) and
(\ref{diagff2}). 
Fitting this they obtain
\beas
\delta&=&(12.23 \pm 1.2)\times 10^{-3} \\
\phi&=&(116.7\pm5.8)^{o}.
\eeas
Rearranging Eq.~(\ref{expon}) using $e^{i\phi}=\cos\phi+i\sin\phi$ results in
\bea
\sqrt{\bar{s}}&=&(m_\rho^2-m_\rho\Gamma_\rho \cot\phi)^{1/2} 
\label{s} \\
y&=&-\frac{\delta\cos\phi[(\bar{s}-m_\rho^2)^2+m_\rho^2\Gamma_\rho^2]}{
am_\rho^2(\bar{s}-m_\rho^2)}.
\eea
This gives $\sqrt{\bar{s}}=792.18$ MeV, close, but not identical, 
to the $\omega$ mass. Substituting $\sqrt{\bar{s}}=m_\omega$
in Eq.~(\ref{s}) reproduces the 
expression for the Orsay phase obtained by Coon {\it et al}. (Eq.~(12) of
Ref.~\cite{CSM}). They also have a contribution to the Orsay phase
{}from a phase difference between the couplings of the vector mesons 
to the photon (as discussed in Sec.~\ref{mixprop}).

The value of $y$ obtained, $(-2.16\pm 0.35)\times10^{-3}$, gives
a value for the \rw mixing parameter
\be
\Pi_{\rho\omega}=-4.225\times10^{-3} {\rm GeV}^2
\ee
which agrees well with the value $-(4.52\pm0.6)\times10^{-3} {\rm GeV}^2$
obtained by Coon and Barrett \cite{CB}, despite the fact that quite
different values for the $\rho$ mass and width are used.
The initial parameterisations of BCP, though, yield a much lower value,
closer to $-(3.7\pm 0.3)\times10^{-3} {\rm GeV}^2$. From this we see that
the value of $\Pi_{\rho\omega}$ is quite sensitive to the parameterisation
of the form-factor.

\subsection{The pion form-factor}
\mb{.5cm}
To make our arguments completely transparent, we shall use the
first form of VMD (as given by Eq.~(\ref{vmd1})) in a calculation of the 
pion form factor \cite{OPTW2}.

For the simplest case of only $\rho$-mesons and pions we would have
{}from Eqs.~(\ref{ffdef}) and (\ref{E1}). 
\be
F_\pi(q^2)=1-\frac{q^2}{g_\rho}\frac{1}{q^2-m_\rho+im_\rho\Gamma_\rho(q^2)}
g_{\rho\pi\pi}.
\label{ourpff}
\ee
We have followed standard assumptions arising from unitarity considerations
\cite{BF}
for the momentum dependence of the $\rho$ width, using the 
form given in Eq.~(\ref{width}) with $\lambda=1$. One could, however, 
simply include
a term of the form $\theta(q^2-4m_\pi^2)$ to the standard
Breit-Wigner imaginary piece, $m_\rho\Gamma_\rho$. This is 
sufficient to model the square root branch point of the pion loop self-energy
at threshold ($q^2=4m_\pi^2$), and to ensure that the imaginary part of
the self-energy is zero below this point. However, in practice we do not
actually show results below threshold.
We take the modern values \cite{HS}
\bea
g_{\rho\pi\pi}^2/4\pi&\sim& 2.9, \\
\label{val1}
g_{\rho}^2/4\pi&\sim& 2.0,
\label{val2}
\eea
coming respectively from $\Gamma(\rho\ra\pi\pi)\sim 149$ MeV and  
$\Gamma(\rho\ra e^+ e^-)\sim 6.8$ MeV. Equating these two constants
actually ruins our fit to data.

To include the contribution of the $\omega$, we shall now use the matrix
element of Eq.~(\ref{diagff2}) determined in Sec.~\ref{mixprop} from 
diagonalising the mixed
propagator. As we are using the first representation of VMD, this will 
provide us with the vector meson contribution to the form factor in the CSV
analogue of the second term on the right hand side of Eq.~(\ref{ourpff}). So, 
including the non-resonant contribution from the direct coupling
of the photon to the pion pair and replacing the Feynman amplitudes appearing
in Eq.~(\ref{nondiag}) or (\ref{diagff2}) with expressions derived from the
VMD1 Lagrangian (Eq.~(\ref{vmdlag})) we have either
\be
  F_\pi = 1 - g_{\rho\pi\pi} \frac{1}{s_\rho} \frac{q^2}{g_{\rho_I}}
  - g_{\rho\pi\pi} \frac{\Pi_{\rho\omega}}{s_\rho s_\omega}
    \frac{q^2}{g_{\omega_I}},  
\label{pure} 
\ee
or
\be
  F_\pi = 1 - g_{\rho\pi\pi} \frac{1}{s_\rho} \frac{q^2}{g_\rho}
  - g_{\rho\pi\pi} \frac{\Pi_{\rho\omega}}{s_\rho-s_\omega}\frac{1}{s_\omega}
    \frac{q^2}{g_\omega},  
\label{real}
\ee
depending on whether one wishes to use the couplings of the pure isospin
states to the photon, as in Eq.~(\ref{pure}), or that of the physical states
to the photon, as in Eq.~(\ref{real}). Use of Eq.~(\ref{pure}) means that
we understand the $\omega\rightarrow\pi^+\pi^-$ decay, before diagonalisation,
as proceeding via the process illustrated in Fig.~\ref{tony}, rather than 
as an $\omega$ which decays exactly like a $\rho$, but modified by a factor
$\Pi_{\rho\omega}/(s_\rho-s_\omega)$, which is the interpretation of 
Eq.~(\ref{real}).

We shall use Eq.~(\ref{real}) to fit to the form-factor data.  The explicit
expression we use is
\be
  F_{\pi}(q^2) = 1 - \frac{q^2 g_{\rho\pi\pi}}{
        g_{\rho} [q^2 - m_\rho^2 + im_\rho \Gamma_\rho(q^2)]} 
      - \frac{q^2 \epsilon g_{\rho\pi\pi}}{
        g_{\omega} [q^2 - m_\omega^2 + im_\omega\Gamma_\omega]}
\label{final}
\ee
where (as in Eq.~(\ref{epseq})),  
\be
  \epsilon = \frac{\Pi_{\rho\omega}}{
        m_\omega^2 - m_\rho^2
        - i(m_\omega\Gamma_\omega - m_\rho\Gamma_\rho(q^2))}
\ee
Since the major decay channel of the $\omega$ is the three pion state, we have
taken the width of the $\omega$ to be a constant \cite{BF}, in contrast to the
case of the $\rho$ which is given by Eq.~(\ref{width}) with $\lambda=1$.  This
approximation is unlikely to seriously affect our results since the width of
the $\omega$ is so much smaller than that of the $\rho$.  We use the Particle
Data Group's (PDG) \cite{pdg94} value of $\Gamma_\omega=8.43$~MeV. For similar
reasons, any momentum dependence in \rw mixing is of little consequence {\em
for the time-like pion form-factor}.  Hence for now we take $\Pi_{\rho\omega}$
to be a constant.  Of course, from the arguments presented in
Sec.~\ref{rwzero} we expect the momentum dependence of $\Pi_{\rho\omega}$ to
be crucial in extrapolations into the space-like region.

It is of some interest now, to compare our form for the form-factor, 
Eq.~(\ref{final}) to that used by D\"{o}nges {\it et al.} \cite{DSM} who also
use the first representation of VMD. In contrast to what we have done, though,
they couple the $\omega$ directly to the pion state (in the same way as the 
$\rho$) and neglect \rw mixing. Their form factor would be equal to ours if
the $g_{\omega\pi\pi}$ used by them where equal to our
$\epsilon g_{\rho\pi\pi}$.
However, because $\epsilon$ is
a complex quantity in general, 
using real numbers, as D\"{o}nges {\it et al.} do for
$g_{\omega\pi\pi}$, is
insufficient. They acknowledge this by stating that ``phases could be
chosen to correctly describe \rw interference."

The coupling of the omega to the photon has long been considered to be 
approximately 1/3
that of the $\rho$ to the photon \cite{GFQ}, and this is supported in a recent
QCD-based investigation \cite{DM}. BCP \cite{BCP} use the leptonic partial
rate \cite{pdg94} to obtain
\bea
\frac{g_\omega}{g_\rho}&=&
\sqrt{\frac{m_\omega\Gamma(\rho\ra e^+e^-)}{m_\rho\Gamma(\omega\ra e^+e^-)}} \\
&=&3.5\pm 0.18.
\label{thisisit}
\eea

\begin{figure}[htb]
  \centering{\
     \epsfig{angle=0,figure=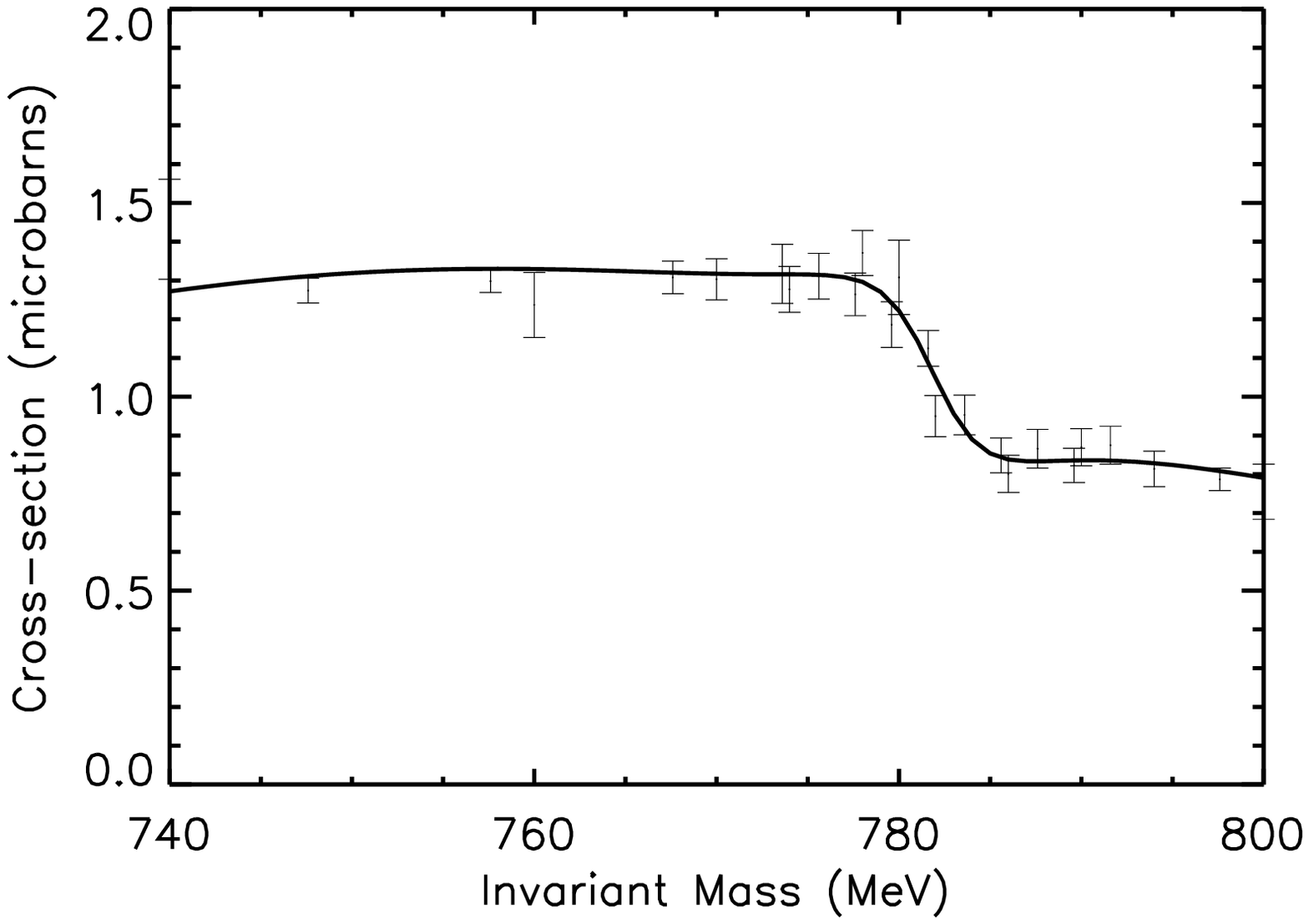,height=10cm}  }
\parbox{130mm}{\caption{Cross-section for $e^+e^-\ra\pi^+\pi^-$ in the region
around the resonance where \rw mixing is most noticeable.}
\label{graph2}}
\end{figure}

With $g_\omega$ fixed in one of these ways, the only remaining free parameter
is the
\rw mixing parameter $\Pi_{\rho\omega}$. It is therefore a simple matter to
fit it to the $e^+e^-\ra\pi^+\pi^-$ cross-section.  The following graphs show
the results of this fit using the form factor of Eq.~(\ref{final}). 
Since the form factor given in Eq.~(\ref{final}) depends only on the ratio
$\Pi_{\rho\omega}/g_\omega$, the choice of
$g_\omega/g_\rho$ significantly alters this.  Using 
the value of 3.5 (Eq.~(\ref{thisisit})) for the ratio  we have, with
$\chi^2/$ d.o.f.=14.1/25,
\be
\Pi_{\rho\omega}=-3800\pm370\;{\rm MeV}^2.
\label{ourval1}
\ee
In this analysis there are two principle sources of error in the value of
$\Pi_{\rho\omega}$. The first is a statistical uncertainty of 310 MeV$^2$
for the fit to data, and the second (200 MeV$^2$) is due to the error
quoted in Eq.~(\ref{thisisit}). These errors are added in quadrature. 
\begin{figure}[htb]
  \centering{\
     \epsfig{angle=0,figure=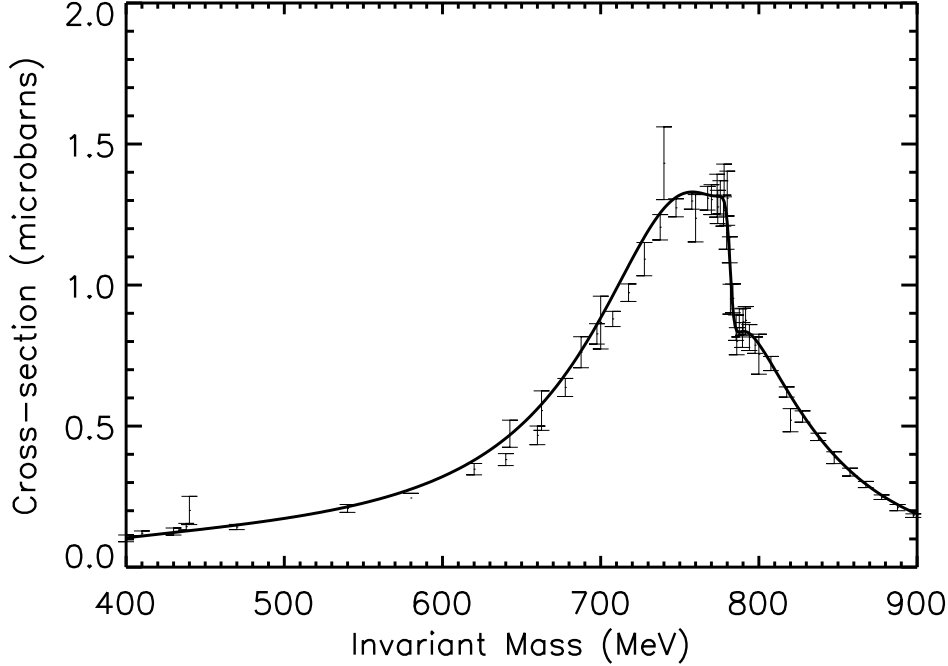,height=10cm}  }
\parbox{130mm}{\caption{Cross-section of $e^+e^-\ra\pi^+\pi^-$ plotted as a 
function of $s^\h$.}
\label{graph1}}
\end{figure}
The result of
our fit to data is shown in Fig.~\ref{graph1} and
resonance region is shown in close-up in Fig.~\ref{graph2}.

It is now of interest to compare our value for $\Pi_{\rho\omega}$ with the 
other values obtained.  Firstly, we observe that
\be
 g_{\omega\pi\pi}  = \frac{\Pi_{\rho\omega}}{
        m_\omega^2 - m_\rho^2
        - i(m_\omega\Gamma_\omega - m_\rho\Gamma_\rho(q^2))}
      g_{\rho\pi\pi}.
\ee
This relation enables us to relate the width for $\omega\ra\pi^+\pi^-$ to the
$\rho$ width via
\be
  \Gamma(\omega\ra\pi\pi)=
     \left| \frac{\Pi_{\rho\omega}}{
        m_\omega^2 - m_\rho^2
        - i(m_\omega\Gamma_\omega - m_\rho\Gamma_\rho(m_\omega^2))}
     \right|^2 \Gamma(\rho\ra\pi\pi) \label{bruce} 
\ee
giving
\be
 \Gamma(\omega\ra\pi\pi)  = 0.157\;{\rm MeV}
\ee
where we have used $\Pi_{\rho\omega}=-3800$~MeV$^2$ corresponding to the
experimental value of 3.5 for the ratio $g_\omega/g_\rho$.  This corresponds
to a branching ratio ${\rm BR}(\omega\ra\pi\pi)=1.86$\%, 
compared with the PDG value of $2.21\pm0.30$.

Comparing Eq.~(\ref{final}) with Eqs.~(\ref{orsayform}), (\ref{beneq}) and
(\ref{frozeq}), we also determine the Orsay phase $\phi$ to be given by
\bea
  \phi &=& \arg\left( \frac{\Pi_{\rho\omega}}{
        m_\omega^2 - m_\rho^2
        - i(m_\omega\Gamma_\omega - m_\rho\Gamma_\rho(m_\omega^2))}
        \right) \nonumber\\
  &=& 101.0^\circ 
\eea
independently of the value of $\Pi_{\rho\omega}$.

\subsection{Previous determinations of $\Pi_{\rho\omega}$}
\mb{.5cm}
It is now of interest to compare our value for $\Pi_{\rho\omega}$ with the 
other values obtained. McNamee {\it et al.} \cite{MSC}
base their predictions on the
decay amplitude of the $\omega$, and obtain
$\Pi_{\rho\omega}$
{}from an approximation to Eq.~(\ref{bruce}),
\be
\Gamma(\omega\ra\pi\pi)=
\left|\frac{\Pi_{\rho\omega}}{im_\rho\Gamma_\rho}\right|^2
\Gamma(\rho\ra\pi\pi).
\label{coon}
\ee
Their answer is thus determined by ${\rm BR}(\omega\ra\pi\pi)$, as is the 
phenomenological plot
by Benayoun {\it et al.} (see their Eqs.~(A.6)-(A.10)), who also take account
(as they have to to plot the cross-section, something not done by Coon and
Barrett) of the relative strengths of the couplings of the mesons to the photon
(appearing in Eq.~(A10) of Ref.~\cite{BF} via $\Gamma_V(e^+e^-)$). Our 
calculation of $F_\pi(q^2)$ does not explicitly feature 
${\rm BR}(\omega\ra\pi\pi)$, and
although all other parameters used by us are completely standard (PDG), 
we obtain a different value for $\Pi_{\rho\omega}$.
The 1974 data gave the result
\be
\Pi_{\rho\omega}=-3400 {\rm MeV}^2,
\ee
which agrees completely with the result we initially obtained 
{}from fitting the form-factor to the cross-section data.

More recently, however, Coon and Barrett repeated this calculation
\cite{CB}, 
using the data from Barkhov {\it et al.} \cite{Bark} which increased
the branching ratio of the $\omega$ decay, ${\rm BR}(\omega\ra\pi\pi)$, from
1.7 to 2.3\% giving 
\be
\Pi_{\rho\omega} = -4520 {\rm MeV}^2.
\label{newcoon}
\ee
We note that while this is the typically quoted value
for the \rw mixing amplitude \cite{b}, it is not the value which provides
the 
optimal fit to the pion EM form-factor.

\subsection{Conclusion}
\mb{.5cm}
We have shown that it is possible to obtain
a very good fit to the pion form factor data
using a $q^2$ dependent coupling for $\gamma^*-\rho$ 
(corresponding to the first representation of VMD), for which $F_\pi(0)=1$ 
irrespective of any values taken by parameters of the model. 

Our extraction of $\Pi_{\rho\omega}$ given by Eq.~(\ref{ourval1}) agrees with
early values, but is in disagreement with the more recent evaluation by other
means (Eq.~(\ref{newcoon})). These differences are essentially a consequence
of the choice of $g_\rho/g_\omega$. With the most recent value of this
ratio, our preferred value for $\Pi_{\rho\omega}(m_\rho^2)$ is 
$-3800\pm 370 {\rm MeV}^2$ (where the error reflects the uncertainty
in $g_\rho/g_\omega$).

\section{Concluding Remarks}
\mb{.5cm}

We have provided a comprehensive review of the ideas of vector meson
dominance with particular application to the pion form-factor. The
less commonly used representation of VMD, which naturally incorporates a
$q^2$-dependent photon$-$vector-meson coupling, seems to be more
appropriate in the modern framework of strong-interactions where quarks
are the fundamental degrees of freedom (we
emphasise that the two representations
of VMD are only equivalent when exact universality is assumed). 
In this context the recent work
suggesting that the \rw mixing amplitude should also vanish at $q^2=0$
in a large class of models is not in the least surprising. A re-analysis
of the pion form-factor using this formulation gave an excellent fit to the 
data, while careful re-analysis near the $\omega$-pole gave a value of
$\Pi_{\rho\omega}(m_\rho^2)=-3800\pm 370 {\rm MeV}^2$. This differs from
other modern fits mainly because we used the most recent value of
$g_\rho/g_\omega$ \cite{pdg94}.

While \rw mixing is of interest in its own right, in the usual framework of
nuclear physics it is vital to our understanding of charge symmetry 
violation. The rapid variation of this mixing amplitude from the 
$\rho$-pole (where it is measured) to the spacelike region (where it
is needed for the nuclear force) completely undermines 
the assumption of $q^2$ independence and generally leads
to a very small CSV potential. One must then look for alternative, 
possibly quark-based, explanations \cite{lb,la,SMG,MSG}.

It has been suggested that a strong $q^2$-variation of the
\rw mixing amplitude would contradict VMD. Our discussion of the original 
formulation of VMD and the corresponding fit to the pion form-factor 
resolves this confusion. 
It has also been suggested that having the photon-$\rho$
coupling go like $q^2$ would be in conflict with data on nuclear shadowing.
By now it is clear that shadowing in photo-nuclear interactions
is appropriately described in the first representation of VMD \cite{IKL}. At
$q^2=0$ the photon decouples from the $\rho$ and interacts directly 
with a nucleon
\cite{SS} to produce a $\rho$ which is then shadowed by its
hadronic interaction.

Of course, it could be argued that no-one has rigorously derived either the
\rw or $\gamma\!-\!\rho$ mixing amplitude from QCD. Until that is done
it is possible to imagine that QCD might generate a contact 
interaction proportional to $\rho_\mu\omega^\mu$. 
However, such an interaction would lead to a constant mixing amplitude as
$|q^2|\ra\infty$, which is a contradiction with the rigorous
result from QCD sum-rules \cite{HHMK,Hat} which show that this amplitude
vanishes in that limit. As the natural
scale in the problem is the vector meson mass it is clear that even in this
case one would expect a substantial variation in $\Pi_{\rho\omega}(q^2)$
between the timelike and space-like regions.

In conclusion we mention some matters needing further work. As the vector 
mesons are not point-like, the mixing amplitudes must deviate from the simple
VMD form eventually. Thus, even the expressions we have given for the
pion form-factor (for example) have a limited range of validity. Finally we
return to the question of the CSV $NN$ force. Although we have argued strongly
against the usual assumptions surrounding the
\rw mixing mechanism, we note that if the $\rho^0NN$ 
(or $\omega NN$ ) vertex
were to have a small CSV (component behaving like 1 rather than $\tau_3$) one
would obtain a similar force (a Yukawa one rather than an exponential one)
\cite{alt1,alt2}.
This deserves further work as do the more ambitious quark-based models
of nuclear CSV.

\vspace{2.5cm}
{\bf Acknowledgments}
\mb{.5cm}

We would like to thank M.~Benayoun, T.~Hatsuda, E.~Henley, 
K.~Maltman, G.A.~Miller and W.~Weise
for their helpful comments at various stages of the work. This work was
supported by the Australian Research Council.


\begin{thebibliography}{99}
\bibitem{We} W.Weise, Hadronic Aspects of Photon-Nucleus Interactions,
Phys. Rep. {\bf 13}, 53--92 (1974).

\bb{Frank} M.R. Frank and P.C. Tandy, Gauge Invariance and the Electromagnetic
Current of Composite Pions, {\it Phys. Rev. C} 49, 478--488 (1994).
\bb{Frank2}M.R.~Frank, Nonperturbative Aspects of the Quark--Photon
Vertex, {\it Phys. Rev. C} 51, 987--998 (1995).
\bb{Frank3} M.R. Frank and C.D. Roberts,
Model Gluon Propagator and Pion and Rho Meson Observables,
{\it Phys. Rev. C} 53, 390-398 (1996).

\bibitem{ba}  W.J.~Marciano, ``The Standard Model", Proceedings of the 1993
Theoretical Advanced Study Institute, Boulder, Colorado, edited by
S.Raby; \\
S.~Godfrey, 
The Standard Model and Beyond,
{\em Physics in Canada} {\bf 50}, 105--113 (1994).

\bibitem{RW} C.D.~Roberts and A.G.~Williams, 
Dyson-Schwinger Equations and their Application to Hadronic Physics,
{\em Prog. in Part. and Nucl. Phys.} {\bf 33}, 477--575 (1994).

\bibitem{GW} D.J.~Gross and F.~Wilczek, Asymptotically Free Gauge Theories,
{\em Phys. Rev. D} {\bf 8}, 3633--3652 (1973).
\bb{Craig} C.D.~Roberts, Electromagnetic Pion Form-Factor And Neutral Pion
          Decay, {\it Nucl. Phys. A} 605, 475-495 (1996).

\bibitem{Nam} Y.~Nambu, Possible Existence of a Heavy Neutral Meson,
{\em Phys. Rev.}  {\bf 106}, 1366--1367 (1957).

\bibitem{CKGZ} G.F.~Chew, R.~Karplus, S.~Gasiorowicz and F.~Zachariasen, 
Electromagnetic Structure of the Nucleon in Local-Field Theory,
{\em Phys. Rev.}  {\bf 110}, 265--276 (1958).

\bibitem{FF} W.R.~Frazer and J.R.~Fulco, 
Effect of a Pion-Pion Scattering Resonance in Nucleon Structure,
{\em Phys. Rev.}  {\bf 2}, 365--368 (1959).

\bibitem{Sak} J.J.~Sakurai, 
Theory of Strong Interactions,
{\em Ann. Phys. (NY)} {\bf 11}, 1--48 (1960).

\bibitem{YM} C.N.Yang and F.Mills, 
Conservation of Isotopic Spin and Isotopic Gauge Invariance,
{\em Phys. Rev.} {\bf 96}, 191--195 (1954).

\bibitem{KLZ} N.M.~Kroll, T.D.~Lee and B.~Zumino, 
Neutral Vector Mesons and the Hadronic Electromagnetic Current,
{\em Phys. Rev.} {\bf 157}, 1376--1399 (1967).

\bibitem{LZ} T.D.~Lee and B.~Zumino,
Field-Current Identities and the Algebra of Fields,
{\em Phys. Rev.} {\bf 163}, 1667--1681 (1967).

\bibitem{CL} T.P.~Cheng and L.F.~Li, {\em Gauge Theory of Elementary
Particle Physics}, Oxford University Press (1984).

\bibitem{Lu} D.~Lurie, {\em Particles and Fields}, John Wiley \& Sons (1968).

\bibitem{Sak2} J.J.~Sakurai, {\it Currents and Mesons}, University
of Chicago Press (1969).

\bibitem{HFN} M.~Herrmann, B.L.~Friman and W.~N\"{o}renberg,
Properties of $\rho$-mesons in Nuclear Matter,
{\em Nucl. Phys.} {\bf A560}, 411--436 (1993).

\bibitem{IZ} C.~Itzykson and J.-B.~Zuber, {\em Quantum Field Theory}, 
McGraw-Hill (1985).

\bibitem{BjD}J.D.~Bjorken and S.D.~Drell, {\em Relativistic Quantum Fields}, 
McGraw-Hill (1965).

\bibitem{BCP} A.~Bernicha, G.~L\'{o}pez~Castro and J.~Pestieau,
Mass and Width of the $\rho^0$ from an $S$-matrix Approach to \ep,
{\em Phys. Rev. D }{\bf 50}, 4454--4461 (1994).


\bibitem{CP} D.G.~Caldi and H.~Pagels, 
Spontaneous Symmetry Breaking and Vector Meson Dominance,
{\em Phys. Rev. D} {\bf 15}, 2668--2676 (1977).

\bibitem{Ban} M.~Bando {\it et al}., 
Is the $\rho$ Meson a Dynamical Gauge Boson of Hidden Local Symmetry?,
{\em Phys. Rev. Lett.} {\bf 54}, 1215--1218 (1985).

\bibitem{Bha} R.K.~Bhaduri, {\em Models of the Nucleon}, Addison-Wesley (1988).

\bibitem{Schecter} J.~Schechter, Electromagnetism
in a Gauged Chiral Model, {\em Phys. Rev. D} {\bf 34}, 868--872.

\bibitem{Hung} P.Q.~Hung, Vector Meson Dominance and the KSRF Relation as 
Consequences of the Interplay between the Standard Electroweak Model and
Hidden Local Symmetry,
{\em Phys. Lett.} {\bf 168B}, 253--258 (1986).

\bibitem{ER} D.~Ebert and H.~Reinhardt, Effective Chiral Hadron Lagrangians
with Anomalies and Skyrme Terms from Quark Flavor Dynamics,
{\em Nucl. Phys.} {\bf B271}, 188 (1986).

\bb{reinhardt} H.~Reinhardt, {\em Prog. Part. Nucl. Phys.} 36, 189




\bibitem{GS} G.J.~Gounaris and J.J.~Sakurai,  
Finite Width Correction to the Vector-Meson-Dominance Prediction For \ep,
{\em Phys. Rev. Lett.} {\bf 21}, 244--247 (1968).

\bibitem{orsay1} J.E.Augustin \etal, 
$\pi^+\pi^-$ Production in $e^+ e^-$ Collisions and \rw Interference.
{\em Nuovo Cimento Lett.} {\bf 2}, 214--219 (1969).

\bibitem{Gla} S.L.~Glashow, 
Is Isotopic Spin a Good Quantum Number for the New Isobars?
{\em Phys. Rev. Lett.} {\bf 7}, 469--470 (1961).

\bibitem{Fub} S.~Fubini, 
Vector Mesons and Possible Violations of Charge-Symmetry in Strong Interactions,
{\em Phys. Rev. Lett.} {\bf 7}, 466--468 (1961).

\bibitem{LS} G.~L\"{u}tjens and J.~Steiner, 
Compilation of Results on the Two-pion Decay of the $\omega$,
{\em Phys. Rev. Lett.} {\bf 12}, 517--521  (1964).

\bibitem{R} F.M.~Renard, \rw Mixing, 
{\em Springer Tracts in Modern Physics}, {\bf 63}, 98--120,
Springer-Verlag (1972). 

\bibitem{GSR} M.~Gourdin, L.~Stodolsky and F.M.~Renard, 
Electromagnetic Mixing of $\rho$ and $\omega$ Mesons,
{\em Phys. Lett.} {\bf 30B}, 347--350 (1969).

\bibitem{CG} S.~Coleman and S.L.~Glashow, 
Departures from the Eightfold Way: Theory of Strong Interaction Breakdown,
{\em Phys. Rev.} {\bf 134}, B671-B681 (1964).

\bibitem{GFQ} A.S.~Goldhaber, G.C.~Fox and C.~Quigg,
Theory of \rw Interference in $\pi^+\pi^-$ Production,
{\em Phys. Lett.} {\bf 30B}, 249--253 
(1969).

\bibitem{SW} R.G.~Sachs, J.F.~Willemsen, 
Two Pion Decay Mode for the $\omega$ and \rw Mixing,
{\em Phys. Rev. D} {\bf 2}, 133--138 (1970).

\bibitem{CS} S.Coleman and H.J.Schnitzer, 
Mixing of Elementary Particles, 
{\em Phys. Rev.} {\bf 134}, B863-B872 (1964).

\bibitem{HS1} J.~Harte and R.G.~Sachs, 
Mixing Effects for $\phi$, $\omega$, and $\rho^0$ mesons,
Phys. Rev. {\bf 135} B459-B466 (1964), namely, 
equation 26.

\bibitem{orsay2} D.~Benaksas, \etal,
$\pi^+\pi^-$ Production by $e^+e^-$ Annihilation in the $\rho$-Energy Range
with the Orsay Storage Ring,
{\em Phys. Lett.} {\bf 39B}, 289--293 (1972).
\bb{MOW} K.~Maltman, H.B.~O'Connell and A.G.~Williams, Analysis of \rw
Mixing in the Pion Form-factor, {\it Phys. Lett. B} 376, 19-24 (1996).
\bibitem{HM} E.~Henley and G.A.~Miller in {\it Mesons in Nuclei}
(eds. M.~Rho and D.H.~Wilkinson) Amsterdam, North Holland (1979)


\bibitem{b} G.A.~Miller, B.M.K.~Nefkens and I.~\u{S}laus, 
Charge-Symmetry, Quarks and Mesons,
{\em Phys. Rep.} {\bf 194}, 1--116 (1990). 


\bibitem{Ger} A.~Gersten, G.L.~Greeniaus, J.A.~Niskanen, A.W.~Thomas, 
S.~Ishikawa and T.~Sasakawa,
Test of Charge Symmetry in Few Body Systems,
{\em Few Body Systems} {\bf 3}, 171--194 (1988).

\bibitem{ONS} K.~Okamoto, 
Coulomb Energy of ${\rm He}^3$ and Possible Charge Asymmetry of Nuclear
Forces, {\em Phys. Lett.} {\bf 11}, 150--153 (1964); \\ J.A.~Nolen, Jr. and
J.P.~Schiffer, Coulomb Energies, Ann. Rev. Nucl. Sci. {\bf 19}, 471--526
(1969).

\bibitem{WIS} Y.~Wu, S.~Ishikawa and T.~Sasakawa,
{\em Phys. Rev. Lett.} {\bf 64}, 1875--1878 (1990).

\bibitem{BI} P.G.~Blunden and M.J.~Iqbal, Contribution of Charge Symmetry
Breaking Forces to Energy Differences in Mirror Nuclei, {\em Phys. Lett.} {\bf
B198} 14--18 (1987).

\bibitem{MO} G.A.~Miller and W.T.H~van~Oers, Charge Independence and Charge
Symmetry, nucl-th/9409013, Chapter for   
{\em Symmetries and Fundamental Interactions
in Nuclei}, eds. E.M. Henley and W. Haxton (World
Scientific) 


\bibitem{Ab} R.~Abegg \etal,
Charge Symmetry Breaking in $np$ elastic scattering at 477 MeV,
{\em Phys. Rev. D} {\bf 39}, 2464 (1989);
Precision Measurement of Charge Symmetry Breaking in $np$ Elastic Scattering
at 347 MeV,
{\em Phys. Rev. Lett.} 75, 1711--1714 (1995).

\bibitem{Vig} S.E.~Vigdor \etal,
Charge Symmetry Breaking in $n$ (polarised) $p$ (polarised) scattering
at 183 MeV,
{\em Phys. Rev. C} {\bf 46}, 410--448 (1992).



\bibitem{MilWil} G.A.~Miller, A.W.~Thomas and A.G.~Williams,
Charge Symmetry Breaking in Neutron-Proton Elastic Scattering,
{\em Phys. Rev. Lett.} {\bf 56}, 2567--2570 (1986); \\
A.G.~Williams, A.W.~Thomas and G.A.~Miller,
Charge Symmetry Breaking in Neutron-Proton Elastic Scattering,
{\em Phys. Rev. C} {\bf 36}, 1956--1967 (1987).

\bibitem{Hol} B.H.~Holzenkamp, K.~Holinde and A.W.~Thomas,
Consistent Evaluation of Charge-Symmetry Breaking Effects in
the Neutron-Proton Interaction,
{\em Phys. Lett. } {\bf B195}, 121--125 (1987).

\bibitem{CSM} S.A.~Coon, M.D.~Scadron and P.C.~Mc~Namee, 
On the Sign of the \rw Mixing Charge Asymmetric $NN$ Potential,
{\em Nucl. Phys.} {\bf A287}, 381--389 (1977).

\bibitem{GHT} T.~Goldman, J.A.~Henderson and A.W.~Thomas, 
A New Perspective on the \rw Contribution to Charge-Symmetry
Violation in the N-N Force,
{\em   Few Body Systems} {\bf 12}   123--132 (1992).

\bibitem{IN} M.J.~Iqbal and J.A.~Niskanen, 
The Effect of Off-shell Variations of \rw Meson Mixing on Charge
Symmetry Breaking Neutron-Proton Scattering,
{\em Phys. Lett. B} {\bf 322}, 7--10 (1994).

\bibitem{lb} K.~Saito and A.W.~Thomas, 
The Nolen-Schiffer Anomaly and Isospin Symmetry Breaking in Nuclear Matter,
{\em Phys. Lett. B} {\bf 335}, 17--23 (1994).

\bibitem{la} E.M.~Henley and G.~Krein, 
Nambu--Jona-Lasinio Model and Charge Independence,
{\em Phys. Rev. Lett.} {\bf 62}  2586--2588 (1989).
\bb{alt1} S. Gardner, C.J. Horowitz and J. Piekarewicz,
Charge Symmetry Breaking Potentials from Isospin Violating
Baryon Coupling Constants, {\it Phys. Rev. Lett.} 75, 2462-2465 (1995).
\bb{alt2}S. Gardner, C.J. Horowitz and J. Piekarewicz,
Isospin Violating Meson-Nucleon Vertices as an Alternate Mechanism of 
Charge Symmetry Breaking, {\it Phys. Rev. C} 53, 1143-1153 (1996).
\bb{alt3} J. Piekarewicz, The Okamoto-Nolen-Shiffer Anomaly without
\rw Mixing, nucl-th/9602010.
\bb{alt4} H.R. Christiansen, L.N. Epele, H. Fanchiotti and C.A. Garcia Canal,
Temperature and Density Effects on the Nucleon Mass Splitting,
{\it Phys. Rev. C} 53, 1911-1916 (1996).
\bb{alt5}A.K. Dutt-Mazumder, B.~Dutta-Roy and A.~Kundu,
Matter Induced \rw Mixing, {\it Phys. Lett.} {\bf B399}, 196-200,1997.

\bibitem{SMG} G.J.~Stephenson,~Jr., K.~Maltman and T.~Goldman, QCD Corrections
to QED and Isospin Breaking in the Baryon Spectrum and Vector Meson Mixing, 
{\em Phys. Rev. D} {\bf 43}, 860--868 (1991)

\bibitem{MSG} K.~Maltman, G.J.~Stephenson,~Jr. and T.~Goldman, Charge Symmetry
Breaking in the $A=3$ System and Electromagnetic Penguins (of the Second
Kind), 
{\em Nucl. Phys.} {\bf A530}, 539--554 (1985).

\bibitem{BC} A.~Bramon and J.~Casulleras, 
New \rw Interference Effects in $J/\Psi\ra\pi^=\pi^-\pi^0\pi^0$ Decays,
{\em Phys. Lett. B} {\bf 173}, 97--101 (1986).

\bibitem{CB} S.A.~Coon and R.C.~Barrett, 
\rw Mixing in Nuclear Charge Asymmetry,
{\em Phys. Rev. C} {\bf 36}, 2189--2194 (1987).

\bibitem{KTW} G.~Krein, A.W.~Thomas and A.G.~Williams, 
Charge-Symmetry Breaking, Rho-Omega Mixing and the Quark Propagator,
{\em Phys. Lett. B} {\bf 317} 293--299 (1993).

\bibitem{MTRC} K.L.~Mitchell, P.C.~Tandy, C.D.~Roberts and R.T.~Cahill, 
Charge Symmetry Breaking Via \rw Mixing from Model Quark-Gluon Dynamics,
{\em Phys. Lett. B} {\bf 335}, 282--288 (1994).

\bibitem{HOC} H.B.~O'Connell, B.C.~Pearce, A.W.~Thomas and A.G.~Williams, 
Constraints on the Momentum Dependence of Rho-Omega Mixing,
{\em Phys. Lett. B} {\bf 336}, 1--5 (1994).

\bibitem{PW} J.~Piekarewicz and A.G.~Williams, 
Momentum Dependence of the \rw Mixing Amplitude in a Hadronic Model,
{\em Phys. Rev. C} {\bf 47}  R2462-R2466 (1993).
\bb{Mitchell}K.L. Mitchell and P.C. Tandy, Pion Loop Contribution to
\rw Mixing and Mass Splitting, 
Kent State preprint KSUCNR-02-96, nucl-th/9607025.
\bb{RF} R. Friedrich and H. Reinhardt, \rw Mixing
and the Pion Electromagnetic Form-Factor in the Nambu--Jona-Lasinio
Model, {\em Nucl. Phys.} {\bf A594} 406--418, (1995).
\bb{chiralp1}K.~Maltman, The Mixed Isospin Vector Current Correlator
   in Chiral Perturbation Theory and QCD Sum Rules,
{\it Phys. Rev. D }53, 2563-2572 (1996).
\bb{chiralp2} R.~Urech, \rw Mixing In Chiral Perturbation Theory,
{\it Phys. Lett. B} 355 (1995) 308.
\bb{chiralp3} K.~Maltman, The Vector Current Correlator
$\langle0|T(V^3_\mu V^8_\nu|0\rangle$ to Two Loops in Chiral
Perturbation Theory, {\it Phys. Rev. D} 53, 2573-2585 (1996).
\bibitem{HHMK} T.~Hatsuda, E.M.~Henley, Th.~Meissner and G.~Krein,
The Off-Shell \rw Mixing in the QCD Sum Rules,
{\em Phys. Rev. C }{\bf 49}, 452--463 (1994).
\bb{sum-rules1}M.J. Iqbal, X.~Jin and D.B. Leinweber,
The Effect Of Off Shell Variations Of Rho Omega Meson
          Mixing Element, {\it Phys. Lett. B} 367, 45-49 (1996).
\bb{sum-rules2}M.J. Iqbal, X.~Jin and D.B. Leinweber,
\rw Mixing via QCD Sum Rules with Finite Mesonic Widths,
nucl-th/9507026 (to appear in {\it Phys. Lett. B}).
\bb{shakin}S. Gao, C.M. Shakin, W. Sun,
Many Body Theory of \rw Mixing, {\it Phys. Rev. C} 53, 1374-1382 (1996).
\bibitem{BLP} V.B.~Berestetskii, E.M.~Lifshitz and L.P.~Pitaevskii, {\em
Quantum Electrodynamics},  Permagon Press (1982).  

\bibitem{spec}S.S.~Schweber, {\em An Introduction to Quantum Field Theory},
Row, Peterson and Company (1961).

\bibitem{BF} M.~Benayoun \etal,
Experimental Evidence for the Box Anomaly.
{\em Zeit. Phys. C} {\bf 58}, 31--53 (1993).

\bibitem{Bark} L.M.~Barkov \etal,
Electromagnetic Pion Form-Factor in the Timelike Region,
{\em Nucl. Phys.} {\bf B256}  365--384 (1985).

\bibitem{D} B.~Dally \etal, Elastic-Scattering Measurement of the Negative-Pion
Radius, Phys. Rev. Lett. {\bf 48}, 375--378 (1982).
\bb{OPTW2} H.B.~OConnell, B.C.~Pearce, A.W.~Thomas and A.G.Williams,
\rw Mixing and the Pion Electromagnetic Form-factor, {\it Phys. Lett. B}
354, 14-19 (1995).
\bibitem{HS} T.~Hakioglu and M.D.~Scadron, 
Vector Meson Dominance, One-loop-order Quark Graphs and the Chiral Limit,
{\em Phys. Rev. D} {\bf 43}, 2439--2442 (1991).

\bibitem{pdg94} Particle Data Group, 
Review of Particle Properties,
{\em Phys. Rev. D} {\bf 50}, 1173--1826 (1994) 

\bibitem{DSM} H.C.~D\"{o}nges, M.~Sch\"{a}fer and U.~Mosel, Micrscopic
Model of the time-like electromagnetic form factor of the nucleon,
{\em Phys. Rev. C51}, 950--968 (1995).

\bibitem{DM} G.~Dillon and G.~Morpurgo, QCD parameterisation of the
$\rho\gamma$, $\omega\gamma$ and $\phi\gamma$ couplings: why $f_{\rho\gamma}:
f_{\omega\gamma}=3:1$ in spite of flavour breaking,
{\em Zeit. Phys. C} {\bf 64}, 467--473 (1994)

\bibitem{MSC} P.C.~McNamee, M.D.~Scadron and S.A.~Coon, Particle Mixing and
Charge Asymmetric Nuclear Forces,
{\em Nucl. Phys.} {\bf A249}, 483--492 (1975).



\bibitem{IKL} B.L.~Ioffe, V.A.~Khoze and L.N.~Lipatov, 
{\em Hard Processes} {\bf 1}, North-Holland (1984).

\bibitem{SS} G.A.~Schuler and T.~Sj\"{o}strand, 
Towards a Complete Description of Higher-Energy Photoproduction,
{\em Nucl. Phys.} {\bf B407}, 539--605 (1993).

\bibitem{Hat} T.~Hatsuda, private communication.


\end{thebibliography}
\end{document}